\documentclass[
 aps,
 pra,
 reprint,
 superscriptaddress,
 amsmath,
 amssymb,
 nobibnotes,
 longbibliography,
 floatfix,
]{revtex4-2}

\usepackage{graphicx}
\usepackage{dcolumn}
\usepackage{array}
\usepackage{multirow}
\usepackage{makecell}
\usepackage{bm}
\usepackage{subcaption}
\usepackage{comment}
\usepackage{xcolor,colortbl}
\usepackage{algorithm}
\usepackage{algpseudocode}
\usepackage{hyperref}
\usepackage{silence}
\usepackage{pdfpages}
\usepackage{pgffor}

\makeatletter
\patchcmd{\@outputpage@head}{\@ifx{\LS@rot\@undefined}{}{\LS@rot}}{}{}{}
\makeatother
\begin{document}

\title{Ion-Trap Chip Architecture Optimized for Implementation of {\\  } Quantum Error-Correcting Code}

\author{Jeonghoon Lee} 
 \affiliation{Department of Computer Science and Engineering, Seoul 08826, Seoul National University}
 \affiliation{NextQuantum, Seoul 08826, Seoul National University}
\author{Hyeongjun Jeon} 
 \affiliation{Department of Computer Science and Engineering, Seoul 08826, Seoul National University}
 \affiliation{NextQuantum, Seoul 08826, Seoul National University}
\author{Taehyun Kim} \email{taehyun@snu.ac.kr}
 \affiliation{Department of Computer Science and Engineering, Seoul 08826, Seoul National University}
 \affiliation{NextQuantum, Seoul 08826, Seoul National University}
 \affiliation{Institute of Applied Physics, Seoul 08826, Seoul National University}
 \affiliation{Inter-university Semiconductor Research Center, Seoul 08826, Seoul National University}
 \affiliation{Institute of Computer Technology, Seoul 08826, Seoul National University}
 \affiliation{Automation and Systems Research Institute, Seoul 08826, Seoul National University}

\date{\today}

\begin{abstract}
We propose a scalable trapped-ion quantum-computing architecture that efficiently incorporates quantum error correction.
The chip design exploits orthogonal qubit connectivity by assigning horizontal trap regions to transversal logical gates and vertical regions to nontransversal gates and syndrome extraction, thereby enabling universal gate operations with minimal ion shuttling and reduced hardware complexity.
Using a dedicated software tool, we evaluate the architecture on several benchmark algorithms and scheduling policies for two-dimensional color code of varying code distance.
Our results demonstrate that increasing the code distance by two reduces the effective logical two-qubit gate error probability by approximately two orders of magnitude, reaching values as low as $10^{-8}$ with the $[[31, 1, 7]]$ color code.
This improvement substantially expands the range of algorithms that can be executed reliably, up to scales of a few thousand logical qubits, depending on the algorithmic structure.
Overall, these findings validate the practicality and scalability of the proposed architecture and its control strategies, highlighting a viable route toward fault-tolerant, trapped-ion quantum computing.
\end{abstract}

\maketitle

\section{Introduction} 

Trapped-ion systems are among the most promising platforms for quantum computing~\cite{Cirac95,Haffner08,Debnath16,Kaushal20,Pino21}, offering advantages such as long coherence times~\cite{Wang21} and high-fidelity operations, including state preparation, readout, and gate operations~\cite{Harty14,Ballance16,Wright19,Clark21,Postler22,An22,Loschnauer25}.
Recently, ion-trap systems have scaled beyond 30 qubits, enabling either all-to-all connectivity within a single chain~\cite{Pogorelov21,Chen24} or arbitrary connectivity via ion shuttling between short chains~\cite{Moses23}.
The latter approach---known as the quantum charge-coupled device (QCCD)---was proposed as a scalable method for building trapped-ion quantum computers~\cite{Kielpinski02}.
By varying the voltages applied to electrodes, ions are physically shuttled between multiple trapping regions, and arbitrary pairs of qubits can be colocated within the interaction region.
Consequently, quantum computation in a QCCD is a sequence of ion shuttling and gate operations within these interaction regions.

Since the QCCD proposal, several studies have explored methods for designing scalable ion-trap chips, for instance, by partitioning trapping regions according to their function (e.g., storage or computation) or by introducing hierarchical structures in the memory regions~\cite{Metodi05,Thaker06,Isailovic08}.
Other research has focused on designing ion-trap systems with an electrode layout to support stabilizer measurements of the surface code~\cite{Lekitsch17}.
Separately, photonic interconnects with optical switches have been proposed as another method for scaling ion-trap systems~\cite{MUSIQC11,Monroe14,Sargaran19}.

Another crucial aspect of large-scale ion-trap systems is the efficient transpilation of an input quantum circuit to the target system's native gates~\cite{Maslov17,Wu21,Durandau23,Kreppel23}.
Several studies focused on small ion-trap systems capable of accommodating fewer than a hundred ions, targeting noisy intermediate-scale quantum (NISQ) applications rather than fault-tolerant regimes~\cite{Preskill18}.
Other studies have also incorporated structural aspects of ion traps to further optimize the transpiled circuits~\cite{Murali19,Murali22,Kurlej24}.

In this paper, we propose a chip layout for shuttling-based trapped-ion quantum computers that enables efficient and fault-tolerant logical gate operations.
By assigning distinct trap regions for transversal and nontransversal gates, the layout leverages the inherent connectivity of a two-dimensional surface trap to naturally interleave error correction steps with gate operations.
To validate the proposed chip layout, a dedicated transpiler and simulator are developed to model the behavior of ions on the proposed chip and run several quantum algorithms with a two-dimensional color code.
The simulation results demonstrate that using a quantum error-correcting (QEC) code on this chip significantly enhances the success rate of these algorithms while maintaining a reasonable runtime cost.
This framework aims to pave the way for more robust and efficient quantum systems capable of handling large-scale circuits and advancing beyond the NISQ regime.

The remainder of the paper is organized as follows.
Section~\ref{section: 2d_color_code} provides a brief overview of the two-dimensional color code, including its logical operations and syndrome extraction procedures.
Section~\ref{section: chip_architecture} introduces the proposed chip layout, detailing its structure, advantages, and the assumptions made for its use.
Section~\ref{section: chip_simulation} describes the dedicated software tool, which encompasses a circuit transpiler, scheduler, and error analyzer.
Simulation results for practical quantum algorithms are presented in Sec.~\ref{section: benchmark}, and we conclude in Sec.~\ref{section: conclusion} with a summary and a discussion of future work.

\section{Preliminaries\label{section: 2d_color_code}} 

In this section, we briefly review the two-dimensional (2D) color code, which is the main QEC code considered in this work.
The 2D color code is a topological QEC code where physical qubits are arranged on a two-dimensional surface~\cite{Bombin06,Fowler11}.
The minimal instance is the $[[7, 1, 3]]$ Steane code, and its code distance can be increased by extending the lattice, provided two conditions are met: each vertex must be connected by exactly three edges (trivalent), and each plaquette must be assigned one of three colors such that no two adjacent plaquettes share the same color (three-colorable).
Each physical qubit is located on a vertex of the lattice, and both $X$- and $Z$- type stabilizer measurements are defined on each plaquette to check the parity of neighboring physical qubits.
While these properties allow for a broad set of logical gates to be implemented transversally, they restrict the choice of regular tilings to the (4.8.8), (6.6.6), and (4.6.12) lattices~\cite{Landahl11}.
For example, the (4.8.8) lattice consists of red squares, blue octagons, and green octagons tiled to satisfy these conditions.

After encoding quantum information, fault-tolerant logical operations are essential for running a quantum circuit.
Transversal gates are a straightforward way to achieve fault tolerance~\cite{Nielsen10}.
Single-qubit transversal gates only require physical single-qubit gates, while two-qubit transversal gates require interactions between physical qubits in different logical qubit.
In the 2D color code with a (4.8.8) lattice, all Clifford gates can be implemented transversally~\cite{Landahl11,Kubica15,Lee25}.
However, since universal quantum computation requires non-Clifford gates~\cite{Nielsen10}, and no QEC code can implement all gates in a universal gate set transversally~\cite{Eastin09}, alternative methods are required.
For instance, $Z$-rotation gates at the third level or higher in the Clifford hierarchy ($Rz(\theta_p), \theta_p=2\pi/2^p$, $p\ge3$) are nontransversal in the 2D color code, and can be implemented by gate teleportation by consuming a magic state $|R_p\rangle_L=(|0\rangle_L+e^{2\pi i/2^p}|1\rangle_L)/\sqrt{2}$~\cite{Zhou00,Campbell16,Mooney21}.
This includes the fault-tolerant $T_L\left[ = Rz(\theta_3) \right]$ gate, which is implemented with the magic state $|R_3\rangle_L=(|0\rangle_L+e^{i\pi /4} |1\rangle_L)/\sqrt{2}$~\cite{Bravyi05}.

To extract information about errors---defined as the error syndrome---in topological QEC codes, including the 2D color code, auxiliary measurement qubits are measured after being entangled with a limited (typically constant) number of physical data qubits.
After measurement and decoding, these qubits can be reused in subsequent rounds of syndrome extraction~\cite{Guillaume10,Katzgraber10,Sahay22,Lee25}.

The core idea behind the proposed chip architecture is the separation of regions for transversal and nontransversal gate operations.
We thus primarily focus on the (4.8.8) 2D color code, which, in addition to its rich set of transversal gates, has the advantage of achieving a given code distance with the minium number of physical qubits among various 2D color code tilings~\cite{Landahl11}.
However, the chip architecture is not strictly limited to this specific code; any Calderbank–Shor–Steane (CSS) code could be adopted, as they feature the transversal \textsc{cnot} gate~\cite{Gottesman97}.

\section{Chip Architecture} 
\label{section: chip_architecture}

\begin{figure}[htbp]
   \includegraphics[width=7.2cm, trim=6 6 9 8]{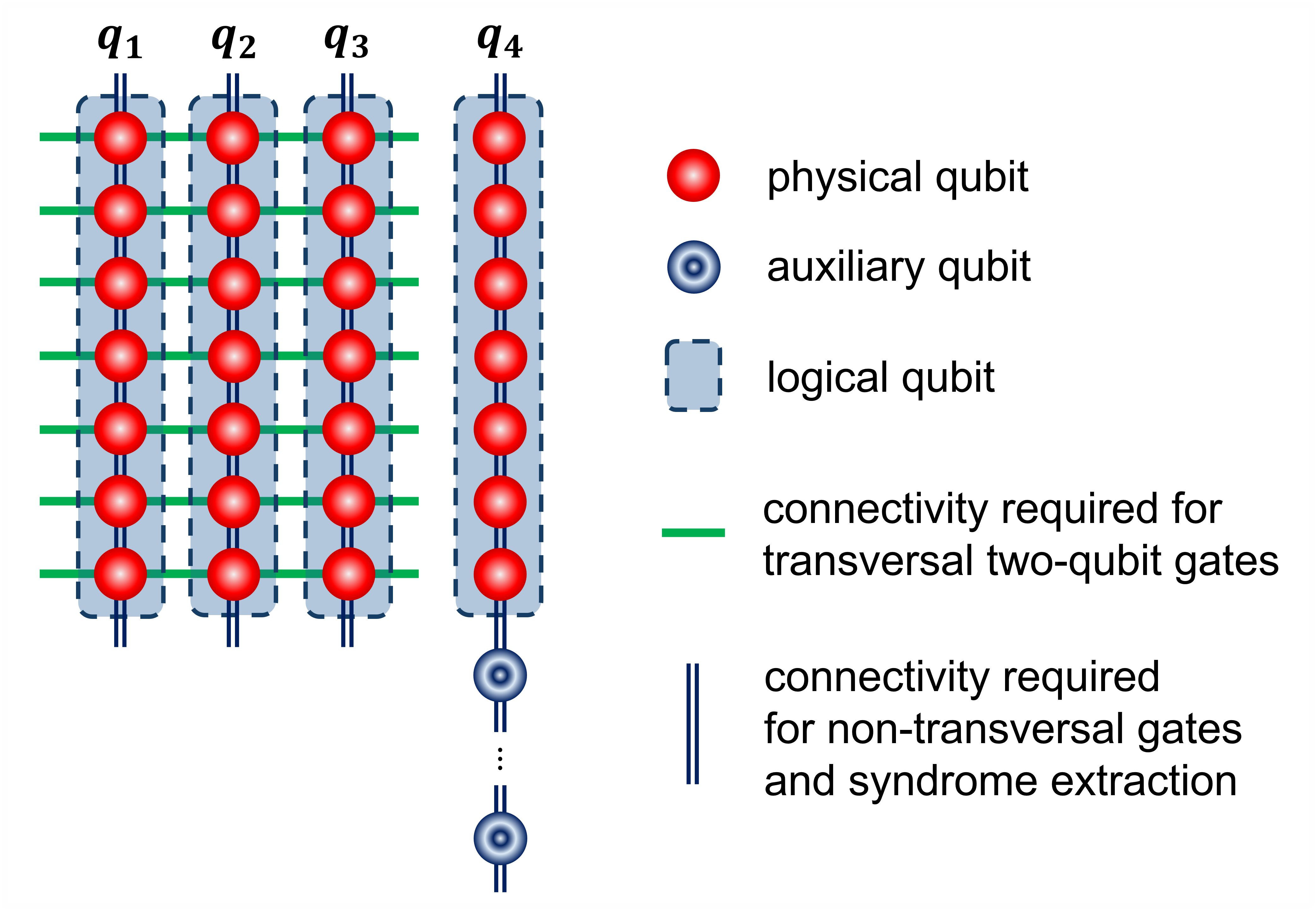}
   \caption{Orthogonal connectivity requirements for transversal versus nontransversal gates and syndrome extraction.
   For the case of the $[[7, 1, 3]]$ code, each logical qubit is encoded using seven physical qubits.
   Transversal two-qubit gates require the qubit connectivity represented by the horizontal lines.
   In contrast, nontransversal gates and syndrome extraction require interactions with auxiliary qubits, represented by the vertical double line.}
   \label{fig: qubit_connectivity}
\end{figure}

A key observation is that two distinct types of qubit connectivity are required to support universal gate operations and syndrome extraction when using the (4.8.8) 2D color code with magic state consumption for nontransversal gates.
As illustrated in Fig.~\ref{fig: qubit_connectivity}, the required connectivity for these operations is effectively orthogonal.
For transversal gates, two-qubit logical gates require interactions only between physical qubits belonging to different logical qubits (horizontal lines); no interaction is required between physical qubits within the same logical qubit, and none is needed for single-qubit logical gates.
In contrast, syndrome extraction and nontransversal gates require interactions between each physical qubit and the auxiliary qubits (vertical double lines).
These auxiliary qubits, which are not used during transversal gate operations, are initialized in the $|0\rangle$ or $|+\rangle$ state for syndrome extraction, or collectively initialized into the state $|R_p\rangle_L$ for nontransversal gates.
To facilitate these interactions, auxiliary qubits are arranged vertically alongside the corresponding logical qubits.
In the example of Fig.~\ref{fig: qubit_connectivity}, logical qubit $q4$ currently utilizes auxiliary qubits, while $q1$ to $q3$ do not, but may make use of them if needed.

\subsection{Chip layout}
\label{subsection: chip_layout}

\begin{figure*}[htbp]
  \includegraphics[width=\textwidth, trim=3 3 3 3]{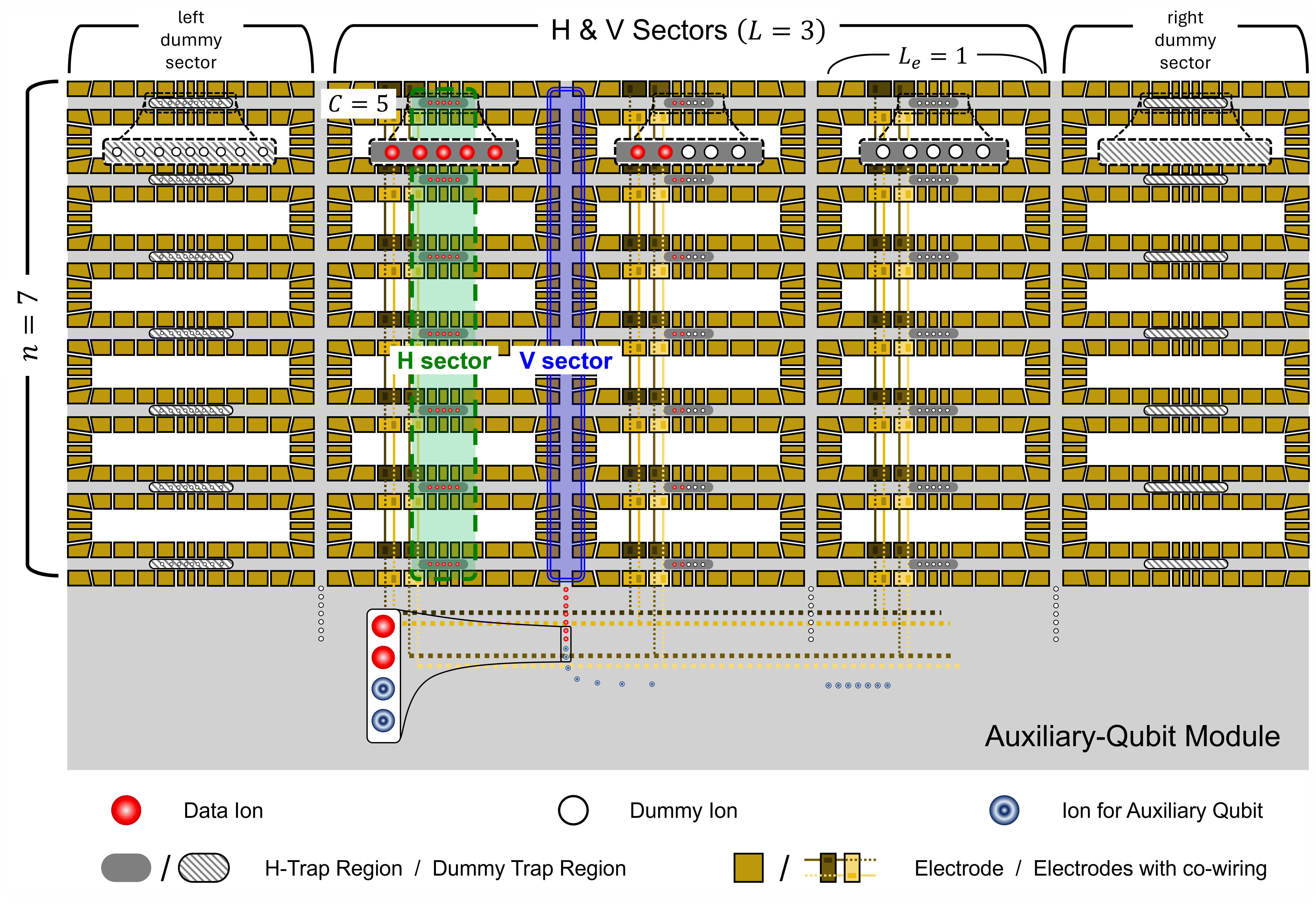}
  \caption{Configuration of the proposed chip layout featuring horizontal trap regions (H-trap regions) and vertical sectors (V sectors) to support diverse qubit-connectivity requirements, with parameters $n=7,C=5, N=8, L=3$, and $L_e=1$.
  $n$ H-trap regions, each containing a chain of $C$ ions, are arranged in parallel to form a horizontal sector (H sector).
  V sectors are placed between H sectors and connected to the auxiliary-qubit module.
  Note that rf electrodes are omitted for clarity, and the figure is not to scale; the actual sizes, numbers, and shapes of components, including ions and electrodes, will differ in practice.
  Sets of electrodes that share a common voltage are co-wired and shown in the same color; for clarity, not all such sets are displayed.}
  \label{fig: chip_layout}
\end{figure*}

Figure~\ref{fig: chip_layout} conceptually depicts the proposed chip layout, designed to support the required connectivity for universal gate operations.
The layout consists of several electrodes forming a grid-like structure where ions can be trapped or shuttled.
Horizontal connectivity is achieved using a horizontal ion chain of length $C$ in a region called a horizontal trap region (H-trap region).
To utilize an $[[n, k, d]]$ QEC code, $n$ parallel H-trap regions are arranged vertically on the chip, and the $n$ physical qubits that encode a single logical qubit are assigned to $n$ vertically aligned ions (called data ions), each in a different H-trap region.
Therefore, a group of $n$ parallel H-trap regions, denoted as an H sector, contain up to $C$ logical qubits.
As H sectors are dedicated to performing transversal gates, this arrangement allows any transversal gate to be executed without shuttling or \textsc{swap} operations among those $C$ qubits.

In contrast, a vertical sector (V sector) lies between two H sectors, contains $n$ X-junctions, and is dedicated to nontransversal gates and syndrome extraction.
To perform these operations, the $n$ data ions constituting a single logical qubit are first shuttled horizontally from one of the adjacent H sectors and then vertically shuttled down through a V sector to a module called the auxiliary-qubit module.
This module can provide auxiliary qubits initialized in the $|0\rangle$ or $|+\rangle$ states to enable syndrome extraction and the magic state $|R_p\rangle_L$ for nontransversal gates, if necessary.
Following these operations, the $n$ data ions are redistributed back into the $n$ H-trap regions of the next adjacent H sector via a V sector.

The entire chip layout is a simple repetition of $L$ H sectors interleaved with $(L-1)$ V sectors.
While each H-trap region contains exactly $C$ ions, some of them are data ions corresponding to physical qubits to encode a logical qubit while others are dummy ions.
These dummy ions act as placeholders to maintain a consistent chain length of $C$ in each H-trap region, which is crucial for uniform operation and control.
To support shuttling operations described later in Sec.~\ref{subsection: rules}, the rightmost $L_e (\ge 1)$ H sectors should be initially filled with dummy ions only.
The system can therefore accommodate at most $(C+1)(L-L_e)$ logical qubits.
Additionally, the chip includes two extra dummy sectors at its far ends to manage dummy ions.
Unlike H sectors, we assume that each dummy trap region in dummy sectors can accommodate arbitrary number of dummy ions, but no meaningful gate operation is performed.

Note that Fig.~\ref{fig: chip_layout} is not drawn to the scale of a real ion-trap chip.
For example, the number and shape of electrodes may vary depending on the experimental setup, ion species, and the value of $C$.
However, the chip layout is based on experimentally demonstrated techniques such as splitting, merging, and shuttling an ion between separated ion chains, as well as turning an ion around X junctions.
Table~\ref{tab: parameters} summarizes the definitions of parameters used throughout the paper.

\begin{table}[htbp]
    \centering
    \begin{tabular}{c|c}
        \hline \hline
        notation & definition  \\ [0.5ex]
        \hline
        \makecell{$n$} & \makecell{number of physical qubits to encode \\
        a single logical qubit} \\ [0.5ex]
        \hline
        \makecell{$d$} & \makecell{code distance \\ ($n=(d+1)^2/2-1$ in (4.8.8) color code)} \\ [0.5ex]
        \hline
        \makecell{$C$} & \makecell{ion chain length \\ in H-trap region ($C\ge2$)} \\ [0.5ex]
        \hline
        \makecell{$N$} & \makecell{circuit width (number of logical qubits \\ used in the quantum circuit)} \\ [0.5ex]
        \hline
        \makecell{$L$} & \makecell{number of H sectors} \\ [0.5ex]
        \hline
        \makecell{$L_e$} & \makecell{number of H sectors traversed during \\ a single shuttling phase $(1 \le L_e < L)$} \\ [0.5ex]
        \hline \hline
    \end{tabular}
    \caption{Definition of parameters used throughout this work.}
    \label{tab: parameters}
\end{table}

\subsection{Rules of control}
\label{subsection: rules}

In this section, operational rules that govern gate operations and ion shuttling on the proposed chip are introduced.
These rules are designed to realize the intended efficiency of the architecture by simplifying hardware control and minimizing shuttling overhead.

\paragraph{Initial configuration}
Initially, every H-trap region in the $L$ H sectors contains an ion chain of length $C+1$.
Given a quantum circuit with $N$ logical qubits, these qubits are first allocated starting from the leftmost H sector, while the intermediate V sectors remain empty.
In other words, $\lfloor N/(C+1) \rfloor$ leftmost H sectors are filled with data ions only, and the interleaving V sectors are unoccupied.
Unless $N$ is a multiple of $C+1$, the $\lceil N/(C+1)\rceil$-th H sector contains data-ion chains of length $r$, where $r=N \text{ mod }{(C+1)}$.
In such cases, $C+1-r$ dummy ions are added to each H-trap region to adjust the chain length to $C+1$.
The remaining H sectors are also filled so that each H-trap region contains $C+1$ dummy ions.
For the two dummy sectors at both ends, each dummy trap region in the left dummy sector is filled with more than $(L_e+1)(C+1)-r$ dummy ions so that every ion chain in the H sector can maintain its length $C$ in subsequent shuttling operations.
In contrast, the right dummy sector is initially empty.

\paragraph{Ion shuttling}
Ion shuttling is required to move logical qubits between H sectors and V sectors.
We assume a simplified, synchronized shuttling model in which all ions undergo alternating phases of rightward and leftward motion, with all ions moved synchronously in each phase.
Also, only one logical qubit composed of $n$ ions can be shuttled between each H sector and its adjacent V sector at a time.
The same constraint applies to a group of $n$ dummy ions.
From the initial configuration, the rightmost ion in each H-trap region is shuttled to its right V sector.
If it were a data ion, it is brought together into the auxiliary-qubit module with other $n-1$ data ions from the same H sector.
Such $n$ data ions form a chain with the required auxiliary qubits and the corresponding logical qubit is subject to nontransversal gate and syndrome extraction.
In contrast, if $n$ rightmost ions of the H sector are dummy ions, they are shuttled down along the V sector to behave identically to other data ions in V sectors, but no meaningful gate operations are conducted.
In this way, all ion chains in H sector have length of $C$.
After several gate operations in both the H sectors and the auxiliary-qubit module, the logical qubit composed of $n$ data ions in the auxiliary-qubit module is shuttled back to the V sector and redistributed to the $n$ corresponding H-trap regions in the right H sector, where each data ion is merged into the local chain, temporarily increasing its length to $C+1$.
Subsequent operation sequence consists of splitting and shuttling $n$ ions at the right end of each H sector to auxiliary-qubit module via a V sector, performing gate operations, and merging them again to temporarily form a chain of length $C+1$.
Figure~\ref{fig: chip_layout} shows the configuration after shuttling one step from the initial configuration.

This shuttling process repeats in a single direction (e.g., to the right in the right-shuttling phase) until each dummy trap region in the right dummy sector is filled with $(L_e+1)(C+1)-r-1$ dummy ions; we refer to this as the right-end configuration.
In this way, every logical qubit passes through the V sector at least $L_e$ times during each one-directional shuttling phase.
The rightmost $L_e$ H sectors ($L_e \ge 1$) are reserved exclusively for dummy ions in the initial configuration to support the above shuttling.
In practice, a small value of $L_e$ (e.g., 2 or 3) is sufficient.
When the system is in the right-end configuration, it switches to a left-shuttling phase, during which all ions move to the left.
In the left-shuttling phase, the $n$ ions in the auxiliary-qubit module are redistributed to the left H sector, and the leftmost $n$ ions in each H sector are shuttled to the adjacent V sector on the left.
This process is repeated until the system reaches the left-end configuration, where the leftmost H sector is full of data ions.
This back-and-forth shuttling ensures that all ions are periodically shuttled between the two ends of the chip, with gate operations interleaved with ion movement.

\begin{figure}[tbp]
    \centering
    \begin{subfigure}[b]{8.4cm}
        \centering
        \caption{Ion configuration at $t$-th time step}
        \includegraphics[width=0.97\linewidth, trim=3 3 3 3]{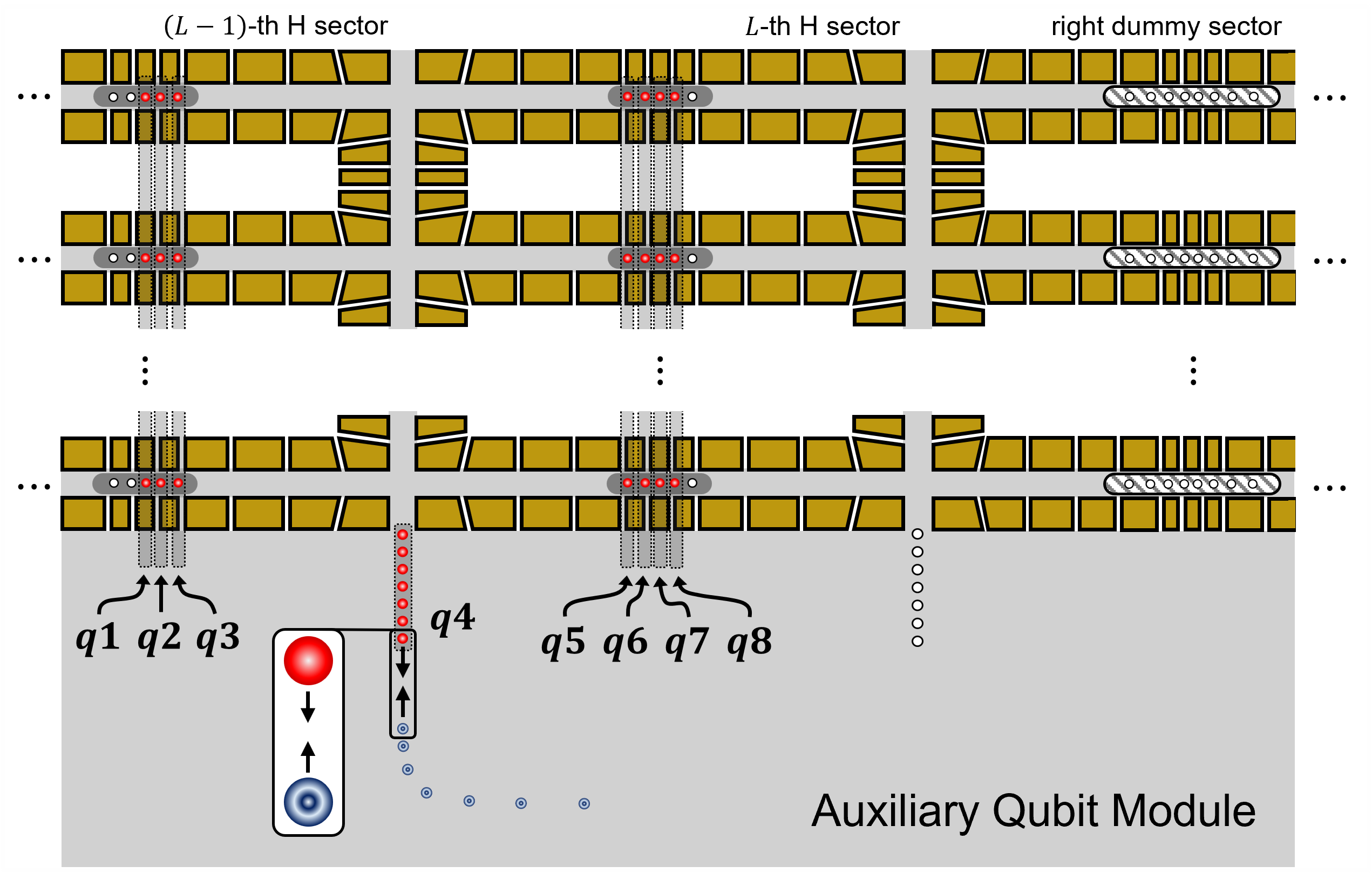}
        \label{fig: shuttle_a}
    \end{subfigure}
    \vspace{0.1cm}
    \begin{subfigure}[b]{8.4cm}
        \centering
        \caption{Ion configuration at $(t+1)$-th time step}
        \includegraphics[width=0.97\linewidth, trim=3 3 3 3]{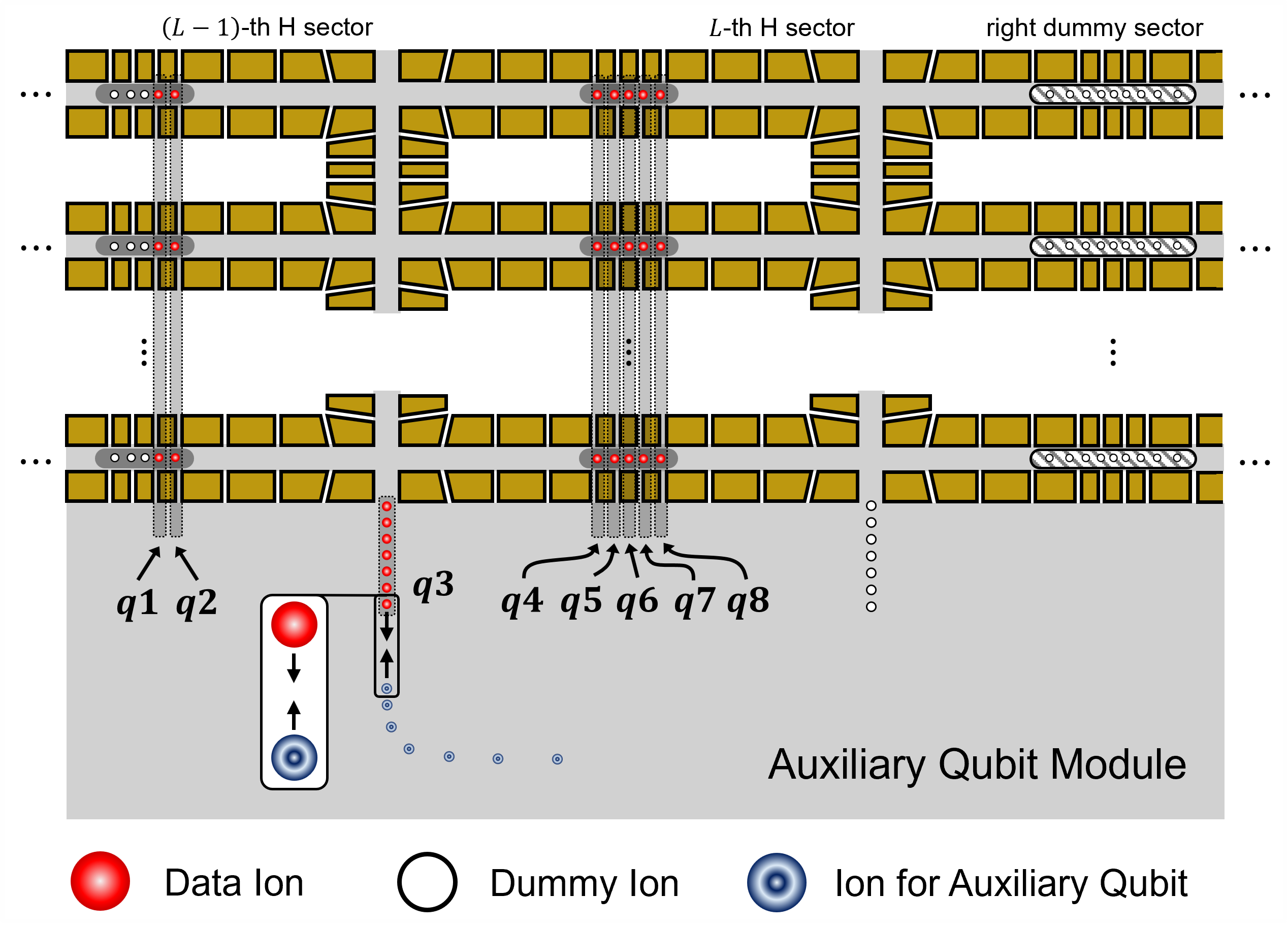}
        \label{fig: shuttle_b}
    \end{subfigure}
    \caption{
    Ion configurations at time step $t$ and $t+1$ during right-shuttling phase.
    The figure is drawn with parameters $n=7,C=5, N=8, L=3$, and $L_e=1$.}
    \label{fig: chip_ion_shuttle_example}
\end{figure}

Figure~\ref{fig: chip_ion_shuttle_example} illustrates an example of a single right-shuttling operation for an 8-qubit circuit with same parameters to Fig.~\ref{fig: chip_layout}.
In Fig.~\ref{fig: shuttle_a}, all logical qubits except $q4$ reside in H sectors, while $q4$ is shuttled to the auxiliary-qubit module via a V sector; this configuration corresponds to time step $t$.
The length of ion chains containing one or more logical qubits is kept at $C=5$ using dummy ions.
After some gate operations, seven data ions of $q4$ return to the V sector and are redistributed across the seven H-trap regions on the right.
Then, the seven data ions of $q3$ are shuttled into the auxiliary-qubit module via the V sector to form an extended ion chain with auxiliary qubits.
Seven dummy ions right next to $q8$ behave identically, but are not merged with auxiliary qubits.
The resulting configuration in Fig.~\ref{fig: shuttle_b} is the right-end configuration.
Therefore, at $(t+2)$-th time step, ions will start to move to the left (i.e., the left-shuttling phase) and repeat until the system reaches the left-end configuration.

\paragraph{Gate operations and syndrome extraction.}
Within each H sector, transversal single- and two-qubit gate operations can be performed without qubit routing.
We conservatively assume that at most one physical operation is performed at any given time within each H-trap region, while multiple operations can be executed in parallel across different H-trap regions.
Although simultaneous multiqubit operations within a single ion chain are technically achievable using multichannel acousto-optic modulators (AOMs)~\cite{Grzesiak20}, such approaches can be limited by practical considerations such as available laser power per channel.
In this work, we deliberately restrict ourselves to a single-operation-per-chain model, consistent with architectures employing sequential beam steering using acousto-optic deflectors (AODs)~\cite{Chen24}.
Importantly, our numerical results in Sec.~\ref{section: benchmark} demonstrate that the proposed architecture still achieves high logical success probabilities under this conservative control assumption, so that multichannel implementations would only further relax timing and resource requirements.
Although our simulations are based on this restricted model, we emphasize that the architecture itself is not tied to any particular gate-implementation technique.
In this way, the system effectively performs one transversal logical gate operation per H sector at any given moment while allowing parallel logical operations across different H sectors.
In contrast, the auxiliary-qubit module, accessed through V sectors, is exclusively reserved for nontransversal gates and syndrome extraction.
If the logical qubit it contains is scheduled for a nontransversal gate, the operation is performed first by consuming a $|R_p\rangle_L$, followed by syndrome extraction.
Otherwise, only syndrome extraction is performed.
All these operations are performed between shuttling operations, and determining which gate to perform first in each sector and when to shuttle all ions is part of the scheduling policy, detailed in Sec.~\ref{subsection: scheduler}.

\paragraph{SWAP operation.}
As shuttling preserves the relative order of logical qubits, two-qubit gates between distant qubits in separate H sectors are not possible without changing their positions.
For example, in Fig.~\ref{fig: chip_ion_shuttle_example}, $q1$ and $q8$ have six logical qubits between them and thus will always remain in different H sectors.
To overcome this, \textsc{swap} gates are used to bring target qubits closer together, analogous to the qubit routing problem in superconducting quantum computers~\cite{Cowtan19}.
However, unlike those systems, an ion-trap system can perform a \textsc{swap} directly between any two qubits within a single H sector, regardless of their physical adjacency.
For a distant two-qubit gate, one target logical qubit is first swapped with the logical qubit at the end of its H sector to reduce the distance to its target pair.
Through a combination of subsequent shuttling and \textsc{swap} operations, two target qubits can eventually be brought into the same H sector.
We assume a logical \textsc{swap} is implemented using three consecutive \textsc{cnot} gates, which is also transversal for the 2D color code and other CSS codes.

\medskip

The primary advantage of this chip layout and its operational rules is the low overhead from ion shuttling.
Since the geometric layout of the chip naturally supports the required qubit connectivity, the design eliminates the need for complex, long-range shuttling and achieves high direct qubit-connectivity while maintaining the scalability of a QCCD approach.
Furthermore, both qubit routing and periodic error correction are naturally integrated into the synchronized shuttling process, ensuring every qubit has an opportunity for correction and nontransversal operations.
Such a synchronized shuttling with ion chains of constant lengths also give advantages in hardware implementation, as detailed in Sec.~\ref{subsection: hardware_cost}.

\subsection{Hardware implementation of chip}
\label{subsection: hardware_cost}

While the proposed chip layout is technically feasible, realizing and controlling the system at large scale requires careful engineering, particularly in electrode wiring and optical control.

The wiring problem--connecting several digital-to-analog converters (DACs) to each electrode--is a critical scalability challenge in ion-trap systems.
Assigning a dedicated DAC to every electrode, while offering maximum flexibility, scales poorly as the number of electrodes grows linearly with the number of qubits.
Following previous approaches that employ co-wiring~\cite{Moses23,Delaney24} or switches~\cite{Malinowski23}, the highly regular geometry and synchronized-shuttling model of the proposed architecture likewise enable comparable hardware efficiencies.

Because all H-trap regions always contain ion chains of identical length, they undergo the same split-shuttle-merge procedure during a shuttling phase.
Namely, the same sequence of waveforms---merging an ion from a V sector into a chain of length $C$ (making the length $C+1$), then splitting another ion from the opposite end of the chain---can be applied to all H-trap regions simultaneously.
This high degree of operational parallelism, depicted as the co-wired set of electrodes in Fig.~\ref{fig: chip_layout}, allows numerous electrodes to share the same set of control lines, keeping the required number of DACs effectively constant, independent of the code distance $n$ or the total number of logical qubits $N$.
The only exceptions are the two dummy sectors at the far ends of the chip, where ion chains of arbitrary lengths must be controlled.
Consequently, there are at most three distinct patterns of ion shuttling.
Similarly, $L$ groups of $n$ ions, shuttled through V sectors, behave identically during vertical shuttling.
When entering a V sector, the ions undergo the same sequence: splitting $n$ ions from the adjacent H-trap regions, vertically shifting them through an X junction, and combining them into a single chain to be delivered to the auxiliary-qubit module.
The difference in merging, depending on whether the pending operation is syndrome extraction or nontransversal gate, is handled entirely within the auxiliary-qubit module.
Therefore, the set of control signals connected to electrodes in each V sector can also be shared across sectors.
Unlike the dynamic electrodes for active ion shuttling, the number of shim electrodes that provide static field compensation necessarily scales with the system size $O(nL)$, although their control overhead can be reduced using multiplexing and switch-based approaches~\cite{Malinowski23,Delaney24}.

Regarding the optical system, the control complexity is independent of the number of physical qubits $n$, as all $n$ ions in a column are driven equally during transversal gate operations.
However, $L$ different sets of optical systems must be controlled independently to support simultaneous gate operations across different H sectors.
While the complexity of nontransversal gates and syndrome extraction operations is relevant to $n$, there is significant potential for optimization since all $O(L)$ V sectors perform only limited, repetitive operations such as stabilizer measurements and magic state consumption via gate teleportation.

\subsection{Other QEC code}
\label{subsection: other_qec_code}

Other than the (4.8.8) 2D color code primarily considered, any CSS codes can also be easily adopted, as they feature the \textsc{cnot} gate as a transversal operation.
The most popular example is the surface code, favored for its nearest-neighbor connectivity and high error threshold~\cite{Fowler09,Fowler12}.
However, using the surface code would be less efficient within our framework because its Hadamard and phase gates are nontransversal, requiring them to be implemented in the V sectors.
A more significant issue is that the surface code is less resource-efficient because the number of physical qubits scales as $n=d^2$ for a code distance $d$, nearly double the $n=(d+1)^2/2 -1$ scaling of the 2D color code~\cite{Bombin07,Landahl11}.
Since the number of H-trap regions is proportional to $n$, adopting the surface code would almost double the hardware cost for the same code distance.
Although the surface code has a higher error threshold, the difference is not dramatic at the moderate code distances considered in this work, and simulations demonstrate that the 2D color code with distance $d \le 7$ effectively suppresses errors in the proposed architecture (see Sec.~\ref{section: benchmark}).
Furthermore, the 2D color code has higher-weight stabilizers [weight-8 in the (4.8.8) color code versus weight-4 in the surface code], but this is easily supported in ion-trap systems for $d \le 7$, due to their high qubit connectivity.
Overall, the (4.8.8) 2D color code is the most suitable QEC code for the proposed architecture.
The application of this framework to other CSS codes, such as the surface code or any other codes, remains an interesting topic for future work.

\section{Chip Simulation} 
\label{section: chip_simulation}

The proposed chip architecture supports a universal quantum gate set, enabling the execution of arbitrary quantum circuits.
In a real-world quantum system, a circuit is first transpiled into an equivalent or approximated circuit composed of native gates supported by the target platform.
The system then executes these gates sequentially or in parallel, along with necessary shuttling and \textsc{swap} operations, according to its scheduling policy.

\begin{figure}[hbt]
    \centering
    \includegraphics[width=0.99\linewidth, trim=4 4 4 4]{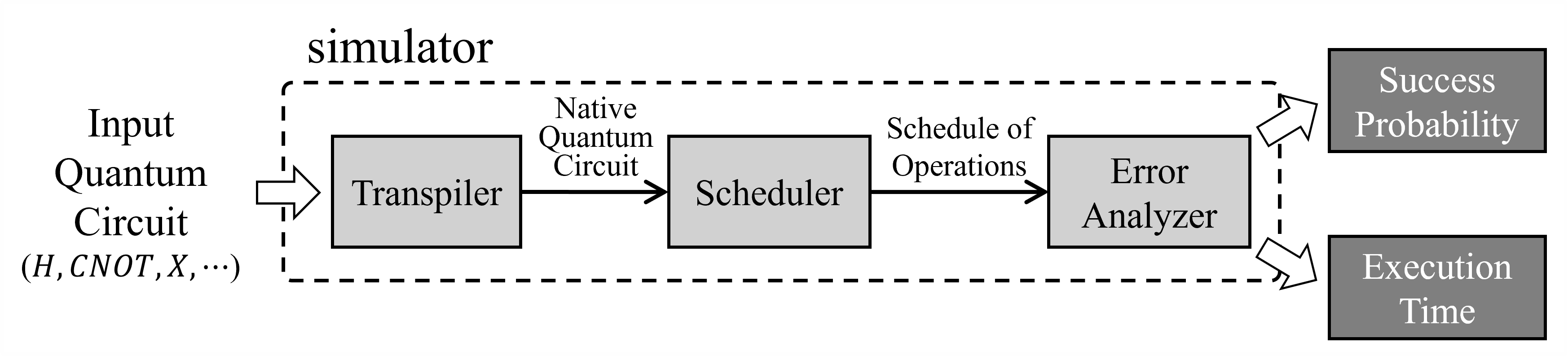}
    \caption{Flow of stages in the software tool, consisting of three main stages: the transpiler, the scheduler, and the error analyzer.}
    \label{fig: simulator_step}
\end{figure}

To evaluate our chip's performance, we developed a dedicated software tool that emulates ion behavior on the proposed architecture.
The tool comprises a transpiler, a scheduler, and an error analyzer that together simulate the entire process from the input circuit to the final output.
It also compares performance against the one without a QEC code, modeled as multiple linearly concatenated ion chains of length $C$.
The flow of the tool is depicted in Fig.~\ref{fig: simulator_step}.

\subsection{Transpiler}
\label{subsection: transpiler}
The transpiler translates an input quantum circuit into the optimized, hardware-specific circuit.
Initially, it performs several optimization passes, including merging adjacent $Z$-rotation gates, reducing the number of Hadamard gates, and canceling consecutive conjugate gate pairs, to reduce gate count and circuit depth~\cite{Barenco95,Nielsen10,Nam18}.
Additionally, some circuits contain multicontrolled Toffoli (\textsc{mcx}) gates, which should be decomposed into \textsc{cnot} and single-qubit gates.
Among several decomposition methods, the V-chain decomposition is used where $n_C-2$ auxiliary qubits are used to decompose an \textsc{mcx} gate with $n_C$ control qubits, resulting in a lowered circuit depth~\cite{Barenco95}.

The circuit is then converted to an equivalent form using only Clifford and $Rz$ gates, where the $Rz$ gates can have arbitrary rotation angles.
For compatibility with the proposed chip architecture, these $Rz$ gates are approximated to a composition of $Rz(\theta_p)$ gates from the higher levels of the Clifford hierarchy.
This ensures that each $Rz(\theta_p)$ gate can be implemented fault-tolerantly in the auxiliary-qubit module.
As a result, the transpiler outputs two circuits: one using this $Rz$ decomposition (for simulation with QEC) and another without it, which is referred as the ``no-QEC" throughout the rest of the paper.

Following these steps, each circuit is converted into a sequence of native gates.
For our simulation, these native gates include the \textsc{gpi} and \textsc{gpi2} gates—single-qubit rotations of $\pi$ and $\pi/2$ around an arbitrary axis on the Bloch sphere equator—and the M\o{}lmer–S\o{}rensen (\textsc{ms}) gate for two-qubit entanglement~\cite{Maslov17,IonqNative24}.
The transpiler applies further optimizations at this native gate level before generating the final circuits for both QEC and no-QEC simulations.

Detailed information about the implementation of the transpiler, including optimization passes and gate conversion rules, is provided in Appendix~\ref{appendix: transpiler}

\subsection{Scheduler}
\label{subsection: scheduler}

\begin{figure}[!hbtp]
    \centering
    \begin{subfigure}[b]{8.5cm}
        \centering
        \caption{Model of the proposed chip layout}
        \includegraphics[width=0.97\linewidth, trim=3 3 3 3]{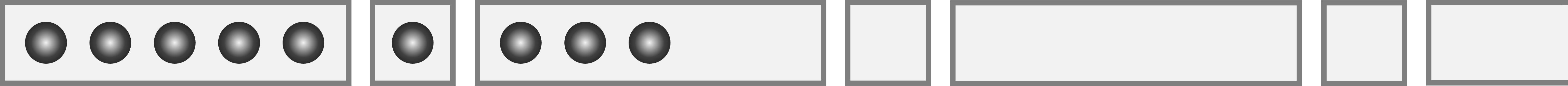}
        \label{fig: chip_model1}
    \end{subfigure}
    \vspace{0.1cm}
    \begin{subfigure}[b]{8.5cm}
        \centering
        \caption{Model of the linear trap chip for no-QEC}
        \includegraphics[width=0.97\linewidth, trim=3 3 3 3]{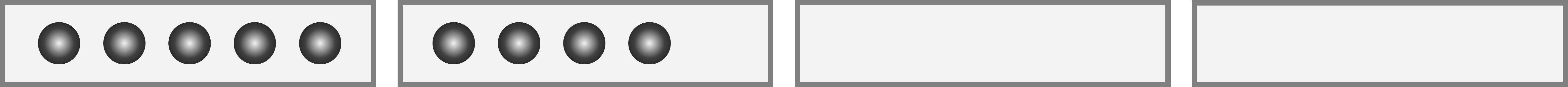}
        \label{fig: chip_model2}
    \end{subfigure}
    \caption{Chip models assumed by the scheduler.
    (a) The model of the proposed chip, where each H sector is represented by a horizontal chain of logical qubits ($C=5$) and each V sector is represented as a single logical qubit.
    (b) The model of linearly concatenated chains for simulation without a QEC code.}
    \label{fig: chip_model}
\end{figure}

The scheduler is responsible for determining which transversal gates to execute, when to shuttle ions, and which qubit pairs to swap within each sector.
To simulate these decisions, the scheduler tracks the configuration and movement of logical qubits on the chip, recording operation timestamps rather than the exact quantum state.
Since all $n$ data ions encoding a logical qubit behave identically, the scheduler models the system by tracking only a single row of logical qubits, excluding dummy ions.
Each H sector is represented as a single horizontal chain of up to $C$ logical qubits, with V sectors modeled as single-qubit chains (see Fig.~\ref{fig: chip_model1}).
To compare the results with the no-QEC case, the corresponding chip layout is modeled as a linear chain of several ion chains of length $C$, as no V sectors are needed without a QEC code (see Fig.~\ref{fig: chip_model2}).

The duration of gate and shuttling operations is critical for scheduling.
While exact times depend on experimental settings, we use relative gate times based on experimentally demonstrated values~\cite{Bermudez17,Bruzewicz19}, as shown in Table~\ref{tab: gate_info}.
The runtimes of single- and two-qubit gates refer to the native gates in the ion-trap system, and runtimes for composite operations like \textsc{swap} and syndrome extraction are calculated from these values.

\begin{table}[ht]
    \centering
    \begin{tabular}{m{3.2cm} m{2.1cm} m{2.1cm}}
        \hline \hline
         \makecell{Gate type} & \makecell{Gate time \\ ratio} & \makecell{Error prob. \\ ratio} \\ [0.3ex]
         \hline
         Two-qubit gate & $5\ (t_{2Q})$ & $1\ (p_{2Q})$ \\ [0.5ex]
         Single-qubit gate & $1\ (t_{1Q})$ & $0.1\ (p_{1Q})$ \\ [0.5ex]
         Measurement & $20$ & $1$ \\ [0.5ex]
         Shuttling operation & $10$ & $0.02$ \\ [0.5ex]
         \hline \hline
    \end{tabular}
    \caption{Ratios of gate times and error probabilities used in all simulations.
    All values are set with respect to the two-qubit gate, based on experimentally demonstrated results~\cite{Bermudez17,Bruzewicz19}.}
    \label{tab: gate_info}
\end{table}
Based on the dedicated models and parameters, scheduling policy actually schedules the gate operations which can largely affect direct qubit connectivity and the frequency of error correction, thereby the circuit's runtime and the quality of its final quantum state as well.
In this work, two heuristic scheduling policies, named unbounded (UB) and vertical trap bounded (VTB) policies, are explored to demonstrate the effectiveness of the proposed chip.

The UB policy aims to execute as many gates as possible in the current configuration, only initiating a shuttling operation when no more gates are executable.
This occurs when all gates are blocked by a pending prerequisite gate, such as a nontransversal gate for a qubit located in an H sector or a two-qubit gate for two qubits in different H sectors.
The policy then strategically inserts \textsc{swap} operations to move the necessary qubits to the ends of the H sectors and shuttle them into the V sectors with subsequent shuttling, repeating this process until the circuit is complete.
In contrast, the VTB policy shuttles all ions as soon as all operations within the V sectors are completed, even if executable gates remain in the H sectors.
In other words, the shuttling interval is determined by the total time required for a nontransversal gate followed by syndrome extraction.
The UB policy prioritizes minimizing shuttling to reduce overall runtime, whereas the VTB policy increases the frequency of error correction, offering each qubit more opportunities to be corrected.
This highlights a fundamental trade-off between circuit runtime and error correction frequency.
More detailed algorithms for these policies are provided in Appendix~\ref{appendix: scheduling_policy}.

\subsection{Error analyzer}
\label{subsection: error_analyzer}

The error analyzer estimates the success rate of a schedule, defined as the probability that a correctable error pattern occurs at every syndrome extraction throughout the circuit execution.
The error analyzer assumes an independent depolarizing error channel and calculates the error probability while scanning the resulting schedule.
When a syndrome extraction is encountered, the analyzer analytically calculates the probability of the correctable error configurations for the (4.8.8) color code at distances $d=3,5,$ and $7$, using a code-capacity model~\cite{Landahl11}.
The probabilities, both for $X$ and $Z$ types, are cumulatively multiplied into the success rate, which is initialized to 1 at the start of the analysis.

Following syndrome extraction, the physical qubit error probability is not reset to zero.
Instead, a circuit-level noise model is used to account for potential logical errors arising from the extraction circuit itself.
This estimates the probability that a syndrome extraction fails, resulting in a logical $X$ or $Z$ flip, by referencing the syndrome extraction circuit of (4.8.8) color code with flagged weight optimization~\cite{Takada24}.
The logical error probabilities for both $X$ and $Z$ are estimated, and the physical qubit error probabilities are reset accordingly.

The error analysis assumes the ratio of gate error probabilities specified in Table~\ref{tab: gate_info}, which are based on experimentally observed values~\cite{Bermudez17,Bruzewicz19}.
The error probabilities for composite circuits are calculated accordingly, and the exact probabilities are specified for each simulation in Sec.~\ref{section: benchmark}.

The runtime of a circuit is defined as the length of the resulting schedule.
Unlike the error analyzer, no separate ``runtime analyzer'' is required, as this value is naturally derived during scheduling.
Since the single-qubit gate is the shortest operation among all operations used, the total runtime is expressed in units of the single-qubit gate time $t_{1Q}$.
Accordingly, the actual execution time of the circuit can be obtained by multiplying the resulting value by the physical duration of a single-qubit gate, typically on the order of a few microseconds.

\section{Performance Analysis with Practical Algorithms} 
\label{section: benchmark}

In this section, we estimate the performance of the proposed ion-trap system through simulations using various benchmark quantum algorithms.
For our analysis, we use a suite of widely recognized quantum algorithms from the Quantum Economic Development Consortium (QED-C), as listed in Table~\ref{tab: benchmark_set}~\cite{Lubinski23, QEDC}.

\begin{table}[h]
\centering
\begin{tabular}{cc}
 \hline \hline
 Gropu 1 & Group 2 \\ [0.5ex]
 \hline \hline
 Phase estimation (PE) & Amplitude estimation (AE) \\ [0.5ex]
 \hline
 Hamiltonian simulation (HS) & Monte Carlo sampling (MC)  \\ [0.5ex]
 \hline
 Quantum Fourier & Variational Quantum \\
 Transform (QFT) & Eigensolver (VQE) \\ [0.5ex]
 \hline
 Bernstein-Vazirani (BV) & Grover search (GS) \\ [0.5ex]
 \hline \hline
\end{tabular}
\caption{Set of benchmark algorithms used in the simulation and their abbreviations.}
\label{tab: benchmark_set}
\end{table}

\begin{figure*}[htbp]
    \centering
    \includegraphics[width=0.98\textwidth, trim=3 3 3 3, clip]{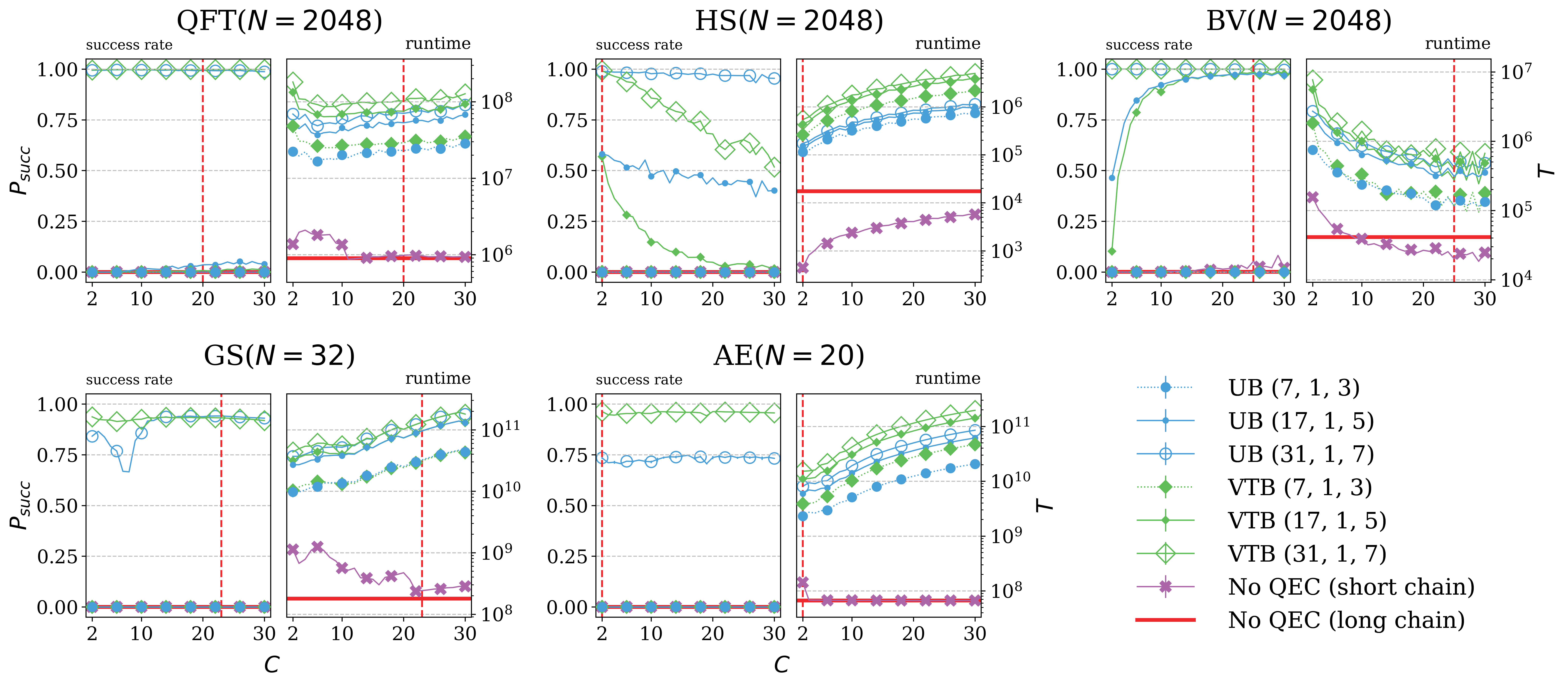}
    \caption{Plot of success rates (left) and runtimes (right) for various algorithms as a function of ion chain length ($C$), assuming a two-qubit gate error probability $p_{2Q}=1 \times 10^{-4}$.
    Other gate error probabilities are scaled according to the ratios specified in Table~\ref{tab: gate_info}.
    The number of logical qubits ($N$) is specified next to each algorithm's name.
    Regardless of the chain length shown on the horizontal axis, ``No QEC (long)" configuration---represented by the markless horizontal line---consistently used a fixed chain length of 100 ions per ion chain ($C=100$).
    The vertical dashed line in each plot denotes the optimal $C$ that yields the highest success rate for each algorithm.}
    \label{fig: trapsize}
\end{figure*}

To begin, we identify the optimal length for the horizontal ion chain for each algorithm by varying its value $C$ within a specific range.
We also assess the success rates at these optimal chain lengths while varying gate error probabilities.
Finally, we introduce and estimate two metrics: the effective error probabilities of gate operations and the expected runtimes for each algorithm.
In brief, the simulations show that the proposed architecture, combined with QEC code, substantially suppresses logical error rates.
Although this introduces additional runtime from error correction and shuttling, the gains in success probability outweigh the overhead, enabling reliable execution of larger circuits with only tens of physical qubits per logical qubit.

\subsection{\label{section: trap_size} Length of horizontal ion chain}

The length of a horizontal ion chain ($C$) directly influences both the success rate and runtime of an algorithm, as it determines the direct qubit connectivity within an H sector and the frequency of syndrome extraction.
A longer chain improves qubit connectivity, thereby reducing the need for \textsc{swap} and shuttling operations.
Conversely, a shorter chain allows for more interleaved V traps, which increases the frequency of syndrome extraction, nontransversal operations, and overall parallelism.
Therefore, the optimal ion chain length involves a trade-off between these two factors.

To find the optimal $C$ for each algorithm, success rates and runtimes are estimated while $C$ is varied from 2 to 30.
Figure~\ref{fig: trapsize} shows the results for representative algorithms with a two-qubit gate error probability of $p_{2Q}=1\times 10^{-4}$.
For algorithms like QFT, HS, and BV, $N=2048$ qubit circuits are chosen, while for other algorithms, whose gate count grows rapidly with $N$, it is selected to make the total gate count comparable to the 2048-qubit QFT circuit.
It is important to note that the $N$ values in the figure exclude auxiliary qubits used for decomposing \textsc{mcx} gates~\cite{Barenco95}.

For comparison, we also simulated two configurations without a QEC code: ``No QEC (short chain)" with variable chain lengths ($C=2$ to $30$) and ``No QEC (long chain)" with a fixed chain length of $C=100$.
Since the no-QEC case does not require V sectors for nontransversal gate, longer chains can provide better connectivity without drawbacks.

The simulation results reveal that the effect of chain length varies significantly across algorithm.
For example, HS performs better with shorter chains because it benefits from increased parallelism and more frequent error correction enabled by small $C$.
In addition, HS circuit in the benchmark set contains relatively short-range interactions, making them less sensitive to the limited long-range connectivity of small $C$.
Conversely, BV is the representative algorithm that benefits from longer chains.
This is because the BV circuit consists only of Clifford gates, which are all transversal in the 2D color code, thereby not being significantly improved with frequent V sector.
Moreover, as most two-qubit gates act on pairs of qubits that are far apart, BV benefits substantially from the long-range connectivity provided by larger $C$.
For other algorithms, the trends are less straightforward, though longer chains generally lead to increased runtime.

In the following sections, all algorithms use the parameter $C$ that yields the highest success rate, which is optimal for each algorithm and $N$.
This optimal $C$, indicated by the vertical dotted lines in Fig.~\ref{fig: trapsize}, is also nearly optimal for runtime, as shorter execution times naturally reduce the probability of errors.

\subsection{Error probability of gate operation}
\label{section: error_prob}

For the simulations in Sec.~\ref{section: trap_size}, the two-qubit gate error probability is set to $p_{2Q}=1\times 10^{-4}$, a value that is demonstrated with the current technology~\cite{Hughes25}.
Moreover, ongoing efforts by several research groups continue to improve gate fidelity in trapped-ion systems.
To determine the physical gate error rates required to achieve high success rates with a reasonable code distance, success rates are estimated while varying the two-qubit gate error probability, with other types of gate error probabilities scaled proportionally according to the ratios in Table~\ref{tab: gate_info}.

\begin{figure}[htbp]
    \centering
    \includegraphics[width=\linewidth, trim=3 3 3 3, clip]{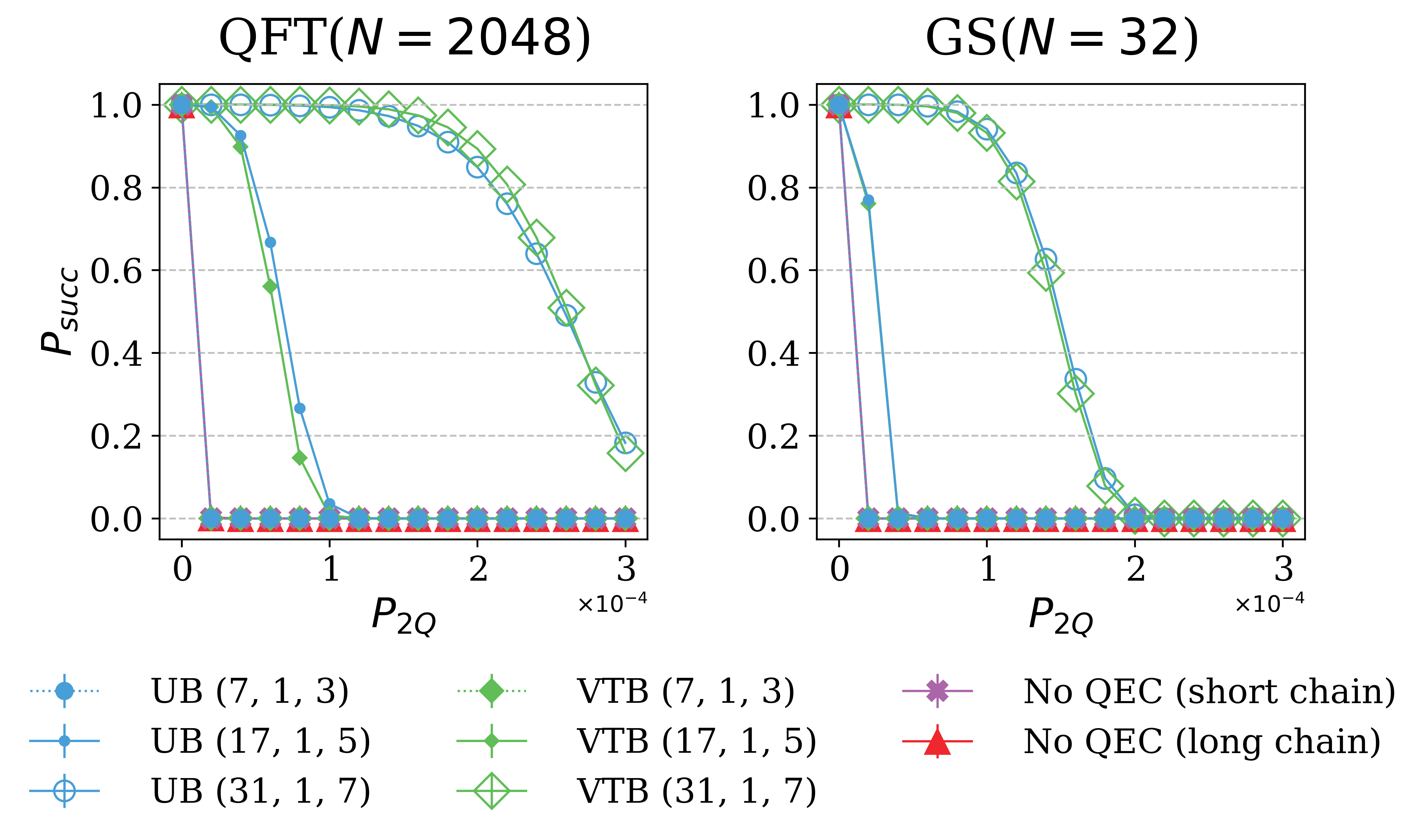}
    \caption{Success rates of 2048-qubit QFT and 32-qubit GS as a function of gate error probabilities.
    The horizontal axis represents the two-qubit gate error probability $p_{2Q}$, with other gate error probabilities scaled proportionally.
    All data points, except for the ``No QEC (long chain)" where $C=100$, assume the optimal $C$ identified in Sec.~\ref{section: trap_size}.}
    \label{fig: error_probability}
\end{figure}

As increasing gate error probabilities always lowers the success rate, the resulting graphs are monotonically decreasing.
We present the results for two representative algorithms: 2048-qubit QFT and 32-qubit GS, as shown in Fig.~\ref{fig: error_probability}.

The simulation results confirm that using a QEC code with a larger distance significantly improves the success rate.
With $p_{2Q}=1\times 10^{-4}$, the $[[31, 1, 7]]$ color code achieves success rates of 99.83\% and 94.10\% for the 2048-qubit QFT and 32-qubit GS, respectively.
In contrast, both algorithms fail in the no-QEC case or when using the smallest 2D color code, the $[[7, 1, 3]]$ code, for all non-zero error probabilities tested.
Although the $[[7, 1, 3]]$ code can correct a single physical-qubit error, it fails to improve the success rate due to the substantial overhead from nontransversal gates and shuttling.
For algorithms such as GS, QFT, and AE (not shown in Fig.~\ref{fig: error_probability}), success rates drop to nearly zero (below $10^{-30}$) in the no-QEC case or with the $[[7, 1, 3]]$ code, even at the smallest physical two-qubit gate error probability tested, $p_{2Q} = 2\times 10^{-5}$.

Although the success rate will ultimately approach zero as the circuit depth increase---regardless of the assumed physical error probability---every algorithm maintains a success rate above 94\% for the circuit sizes tested when assuming a two-qubit gate probability of $1\times 10^{-4}$~\cite{Hughes25}.
Accordingly, subsequent simulations adopt $p_{2Q}=1\times 10^{-4}$ as in Sec.~\ref{section: trap_size}.

\subsection{Number of logical qubits}
\label{section: num_qubit}

\begin{figure}[ht]
    \centering
    \begin{subfigure}[b]{0.454\linewidth}
        \centering
        \caption{Group 1, QEC}
        \includegraphics[width=\linewidth, trim=5 5 5 5, clip]{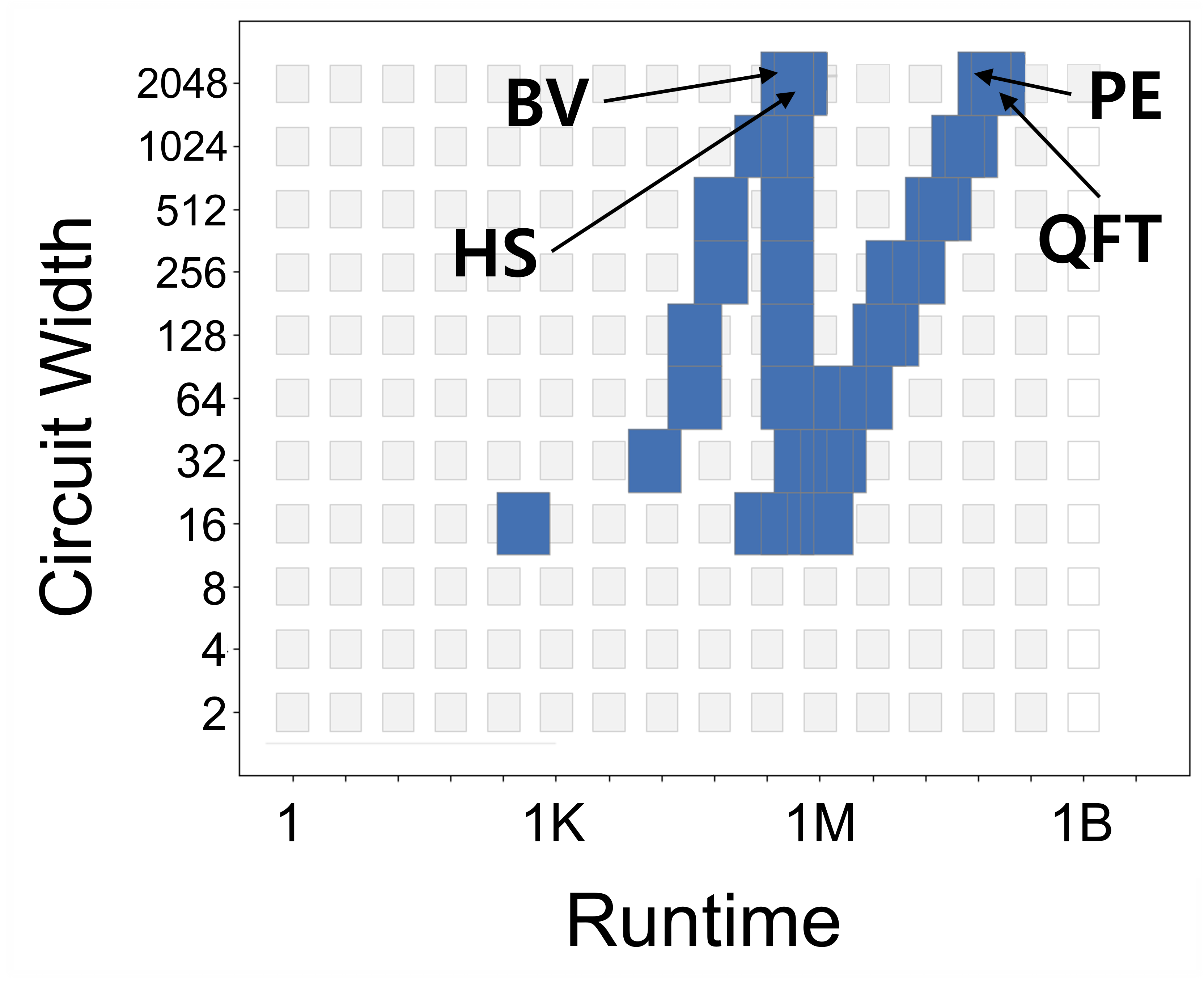}
        \label{fig: volumetric_poly_qec}
    \end{subfigure}
    \hspace{0.01cm}
    \begin{subfigure}[b]{0.518\linewidth}
        \centering
        \caption{Group 1, No-QEC}
        \includegraphics[width=\linewidth, trim=5 5 5 5, clip]{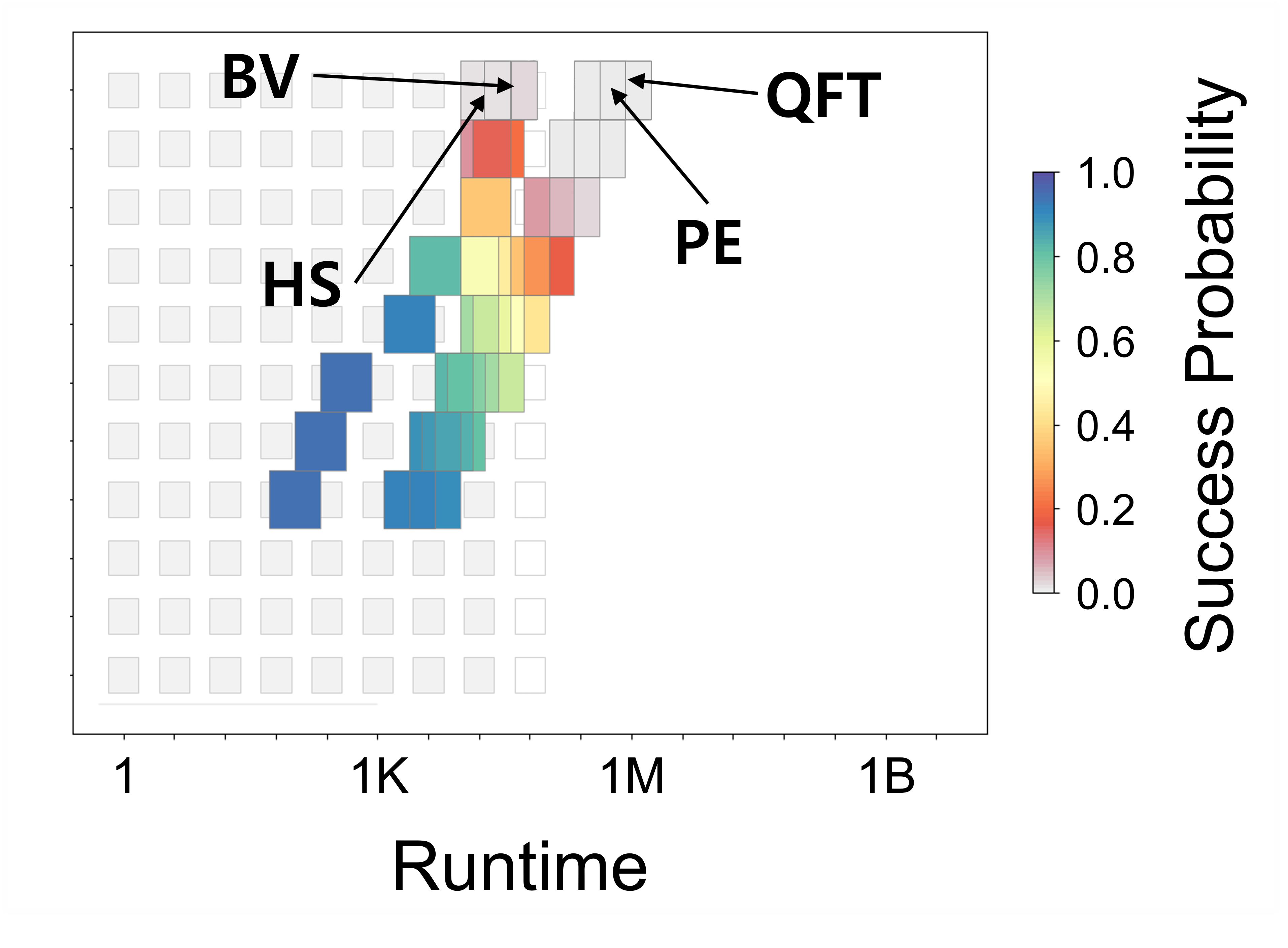}
        \label{fig: volumetric_poly_noqec}
    \end{subfigure}
    \begin{subfigure}[b]{0.443\linewidth}
        \centering
        \caption{Group 2, QEC}
        \includegraphics[width=\linewidth, trim=5 5 5 5, clip]{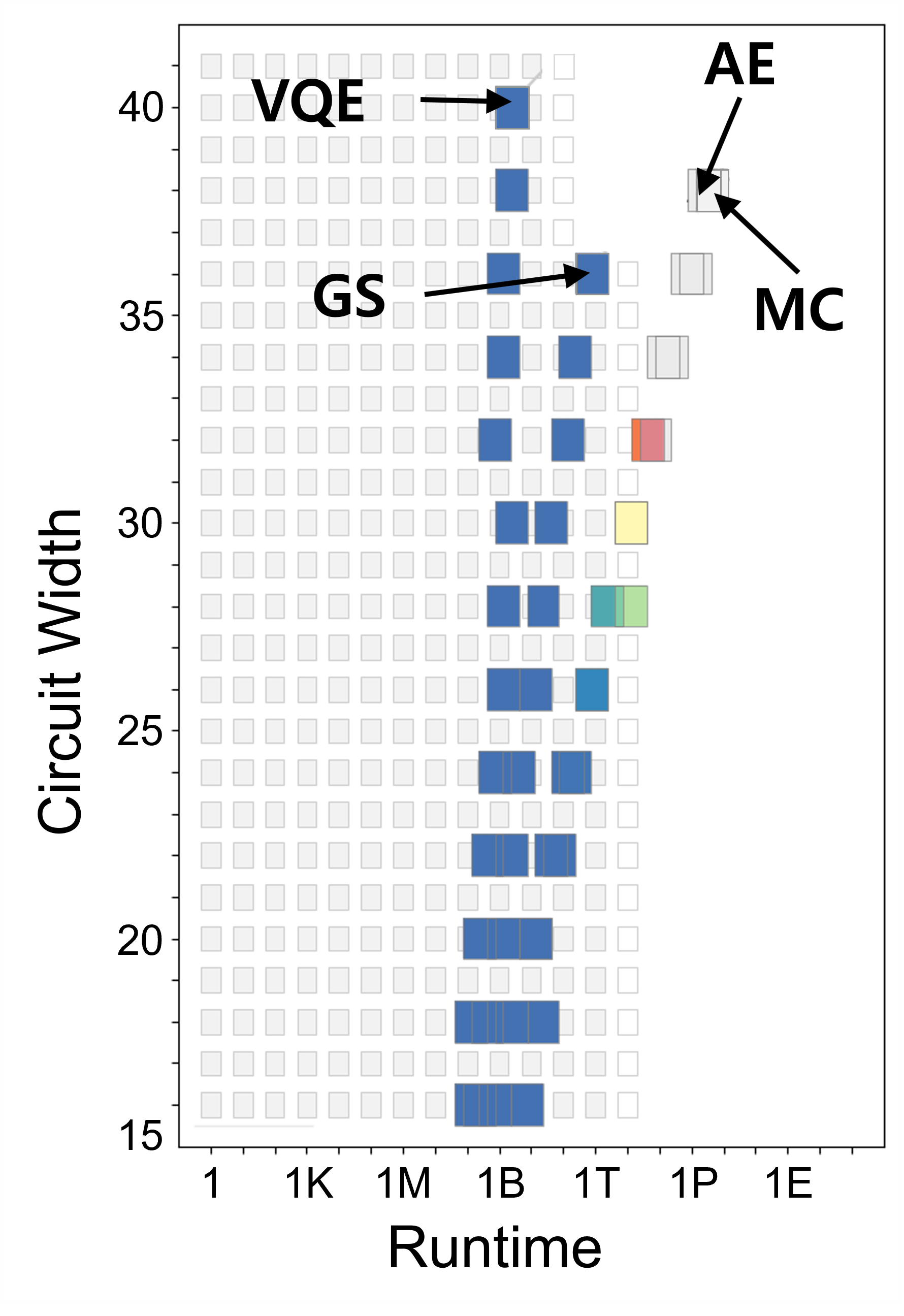}
        \label{fig: volumetric_exp_qec}
    \end{subfigure}
    \hspace{0.05cm}
    \begin{subfigure}[b]{0.525\linewidth}
        \centering
        \caption{Group 2, No-QEC}
        \includegraphics[width=\linewidth, trim=5 5 5 5, clip]{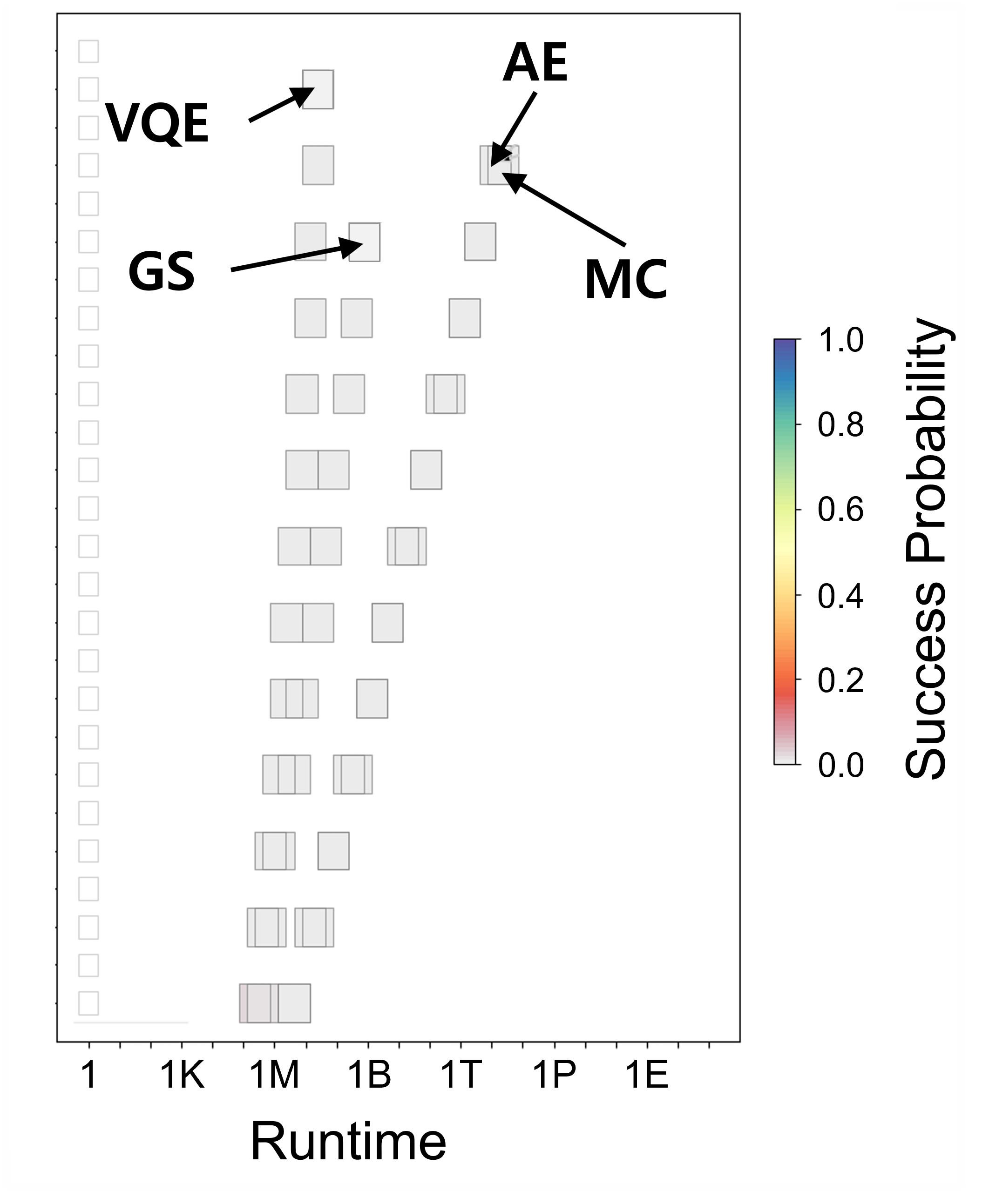}
        \label{fig: volumetric_exp_noqec}
    \end{subfigure}
    \caption{Volumetric benchmarking plots of two groups of algorithms, with QEC and without QEC (no-QEC).
    Eight benchmark algorithms are categorized into two groups based on the scaling of their circuit depth, and each group includes two plots: results using the $[[31, 1, 7]]$ code under the VTB scheduling policy with optimal $C$, and the results for the no-QEC case using a long chain ($C=100$).
    All plots assume a physical two-qubit gate error rate of $p_{2Q}=1\times 10^{-4}$.
    (a), (b) Results for QFT, BV, PE, and HS.
    (c), (d) Results for AE, GS, MC, and VQE.}
    \label{fig: volumetric}
\end{figure}

To demonstrate an advantage over classical approaches, a sufficiently large number of logical qubits is required.  
This section first visualizes the performance improvements of the proposed chip as the number of logical qubits varies, using volumetric benchmarks to illustrate the maximum circuit sizes it can successfully support, as shown in Fig.~\ref{fig: volumetric}.
The volumetric benchmark plots success rates on a two-dimensional space defined by circuit width (number of qubits) and depth~\cite{BlumeKohout20,Lubinski23}.
Due to the large variability in circuit sizes among the benchmarks, the eight QED-C algorithms are grouped according to the complexity of their circuit depth.
For simulations with the QEC code, results from the VTB scheduling policy with the $[[31, 1, 7]]$ code are used; for the no-QEC case, the long chain with $C=100$ is used, as these configurations typically yield the best results in each setting.

The plot clearly reveals a significant performance gap.
Algorithms like QFT and BV, which have a polynomial scaling of circuit depth, achieve high success rates for 2048 qubits with the $[[31, 1, 7]]$ code, specifically over 99.9\% for all 2048-qubit algorithms (see Fig.~\ref{fig: volumetric_poly_qec}).
In contrast, in the no-QEC case, all four algorithms show success rates lower than 50\% even with 512 qubits (see Fig.~\ref{fig: volumetric_poly_noqec}).
For algorithms with exponentially increasing circuit depth, such as AE and GS, the $[[31, 1, 7]]$ code maintains success rates above 90\% for up to 26 logical qubits (see Fig.~\ref{fig: volumetric_exp_qec}), whereas the no-QEC cases almost fail with fewer than 16 qubits (see Fig.~\ref{fig: volumetric_exp_noqec}).
Overall, the proposed architecture substantially extends the maximum circuit depth that can be executed successfully.

\begin{figure*}[hbtp]
    \centering
    \includegraphics[width=0.98\textwidth, trim=5 5 5 5, clip]{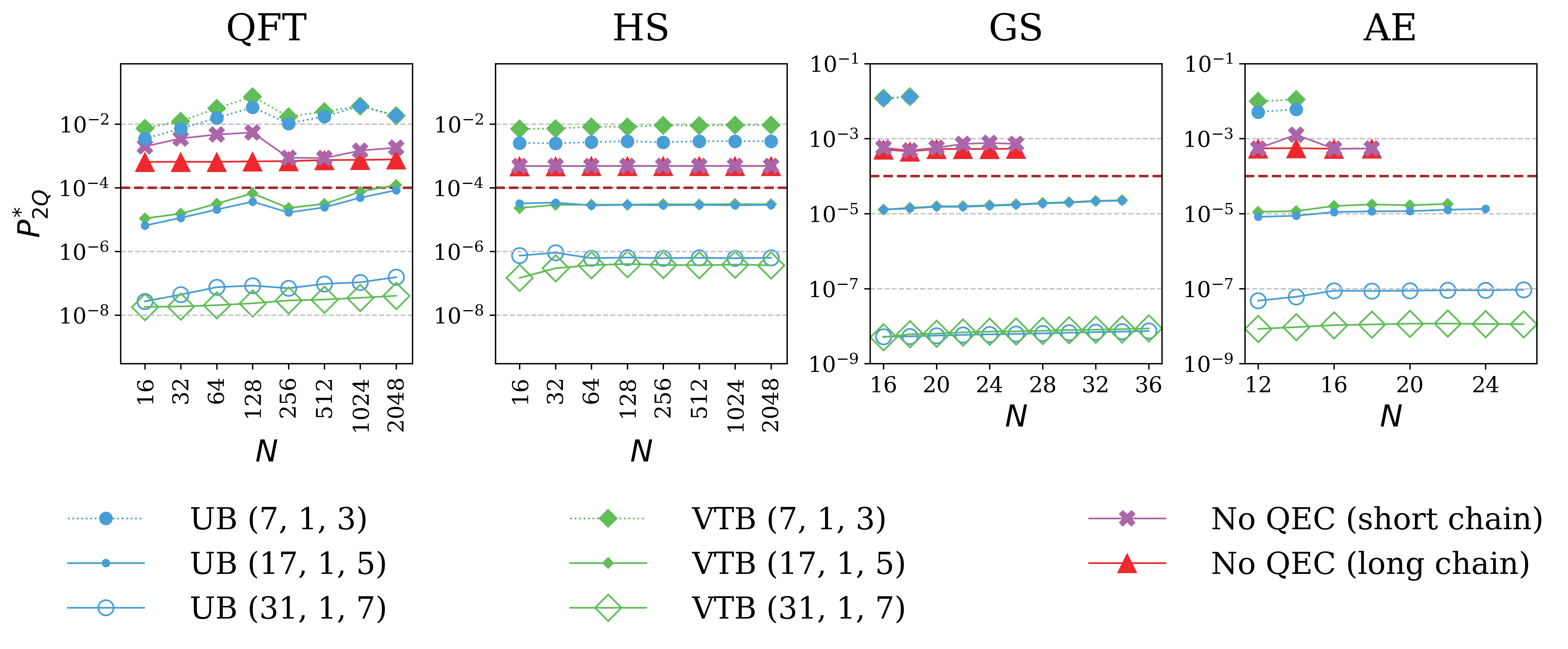}
    \caption{Effective two-qubit gate error probabilities $p_{2Q}^{*}$, calculated using Eq.~(\ref{eq: effective_err_prob}).
    The horizontal dashed line indicates the assumed two-qubit gate error probability, $p_{2Q} = 1 \times 10^{-4}$.
    Data points are omitted for cases where $P_{succ} = 0$, corresponding to $p_{2Q}^{*} = 1.0$.}
    \label{fig: effective_err_prob}
\end{figure*}

Based on the same experimental results, the performance of the proposed architecture is characterized using two metrics: the effective logical gate error probabilities and the expected runtime.
The effective error probabilities of the single-qubit gate $p_{1Q}^{*}$ and the two-qubit gate $p_{2Q}^{*}$ are defined as the logical error probabilities that would reproduce the observed success rate $P_{succ}$ if the benefits of error correction and the overheads from shuttling and $Rz$ gate decomposition were absent.
While the value of $P_{succ}$ is obtained using the depolarizing error channel, the quantities $p_{1Q}^{*}$ and $p_{2Q}^{*}$ are estimated in reverse using the simple multiplicative error-accumulation model of Eq.~(\ref{eq: effective_err_prob}), in which $n_{1Q}$ and $n_{2Q}$ denote the number of single- and two-qubit gate operations, respectively.

\begin{equation}
    P_{succ} = \left(1-p_{1Q}^{*}\right)^{n_{1Q}} \left(1-p_{2Q}^{*}\right)^{n_{2Q}}
    \label{eq: effective_err_prob} 
\end{equation}
It is important to note that $n_{1Q}$ and $n_{2Q}$ are counted from the circuits in the no-QEC case.
This ensures a fair comparison by preventing $p_{2Q}^{*}$ from being overestimated due to the larger $n_{1Q}$ that arises from $Rz$ decomposition in the QEC circuits.
Based on the ratios in Table~\ref{tab: gate_info}, $p_{1Q}^{*}$ is set to $p_{2Q}^{*}/10$, and the corresponding $p_{2Q}^{*}$ is numerically found.
Figure~\ref{fig: effective_err_prob} shows the results of $p_{2Q}^{*}$ for four representative algorithms as a function of the number of logical qubits.
Assumed $p_{2Q}=1\times 10^{-4}$ is marked as the horizontal dashed line in each plot, and data points are omitted for cases where the success rate is zero ($P_{succ}=0$), as this indicates that all circuit executions resulted in failure.

\begin{figure*}[hbtp]
    \centering
    \includegraphics[width=0.98\textwidth, trim=5 5 5 5, clip]{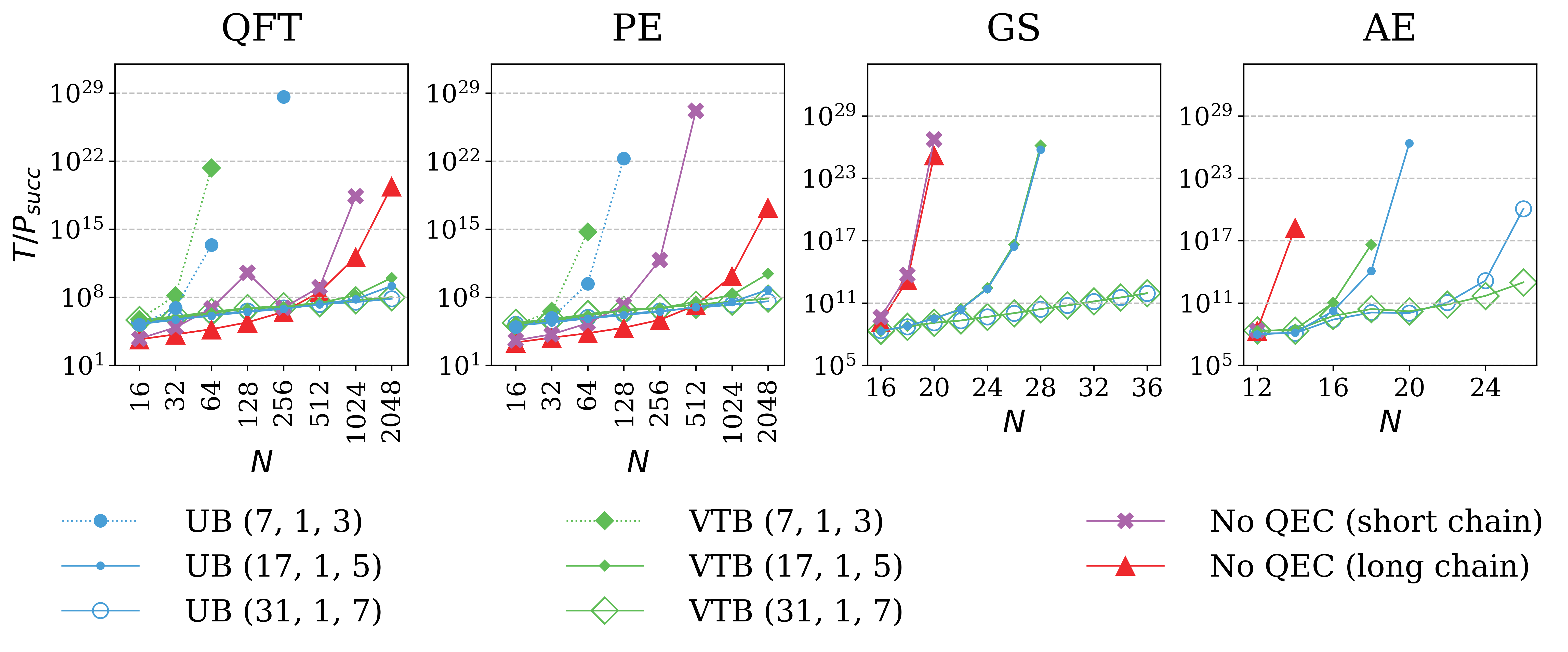}
    \caption{Plot of $T/P_{succ}$, where $T$ denotes the runtime and $P_{succ}$ the success probability of each algorithm with $p_{2Q}=1\times 10^{-4}$.
    The metric represents the expected runtime required for successful circuit completion, with lower values indicating better performance.
    Data points are omitted for cases where $P_{succ}=0$, as this implies $T/P_{succ}=\infty$, indicating that successful execution is not achievable.}
    \label{fig: t_over_p}
\end{figure*}

The results in Fig.~\ref{fig: effective_err_prob} show that the effective error probabilities remain nearly constant across different circuit sizes, demonstrating that the chosen metric properly characterizes the efficiency of error correction and scheduling.
Most importantly, an increase of two in the code distance significantly reduces $p_{2Q}^{*}$: the $[[31, 1, 7]]$ code achieves an effective probability on the order of $10^{-7}$ to $10^{-8}$, while the $[[17, 1, 5]]$ and $[[7, 1, 3]]$ codes yield values around $10^{-5}$ and $10^{-2}$, respectively.
The results of the no-QEC case perform slightly better than those of the smallest code across all algorithms, as the benefit of error correction is insufficient to compensate for the overhead of additional operations.

Overall, increasing the code distance by two effectively reduces the logical gate error probability by more than two orders of magnitude, allowing the distance-five code to outperform the no-QEC baseline.
These results provide numerical evidence for the effectiveness of the proposed chip layout and scheduling strategies in realizing QEC codes and suggest a clear path for scaling logical qubits in ion-trap systems toward the fault-tolerant quantum computation regime.

While the QEC code enhances success rates, it also increases runtime due to overheads such as $d$-times repeated stabilizer measurements and a greater number of gates from $Rz$ decomposition.
To assess this trade-off, the metric, defined as $T/P_{succ}$ where $T$ is runtime, is estimated.
This metric represents the expected runtime to successfully complete an algorithm.
Figure~\ref{fig: t_over_p} shows the results for four representative algorithms, with data points omitted when $P_{succ}=0$ or $T/P_{succ}>10^{30}$.

For most algorithms, the results show that spending additional time on error correction is worthwhile.
For instance, in GS, the values of $T/P_{succ}$ for the $[[17, 1, 5]]$ and $[[31, 1, 7]]$ codes are substantially lower than for the no-QEC case, indicating better performance.
Specifically, the maximum $T/P_{succ}$ observed across all eight algorithms tested with the $[[31, 1, 7]]$ code is $1.42 \times 10^{12}$ for a 26-qubit MC (not shown in Fig.~\ref{fig: t_over_p}).
In contrast, simulations of the no-QEC case often lead to $T/P_{succ}$ rapidly exceeding $10^{100}$ and diverging toward infinity as circuit size grows.
This demonstrates that even small-distance QEC codes significantly improve success rate, outweighing the runtime overhead.

Interestingly, while the VTB policy generally achieves higher success rates, the UB policy often yields better $T/P_{succ}$ values.
This occurs because the more frequent ion shuttling required by the VTB policy provides more opportunities for error correction, but also increases the runtime.
For similar reasons, there are cases in which a smaller code distance---or even the no-QEC case---outperforms larger codes in terms of $T/P_{succ}$ when $N$ is not large.
In these cases, the circuits are not deep enough for the improvement in success rate to offset the runtime overhead.
However, as the circuit size increases, crossover points arise where the benefit of a larger QEC code starts to outweigh its overhead, as observed for QFT and PE.

Since the runtime $T$ is measured as the length of the schedule in units of $t_{1Q}$, the absolute values of $T/P_{succ}$ can be interpreted accordingly.
Assuming $t_{1Q}=10\ \mu s$, a value of $T/P_{succ}=7.10\times10^{7}$ for the 2048-qubit QFT with the $[[31, 1, 7]]$ code under the UB policy corresponds to approximately 12 minutes of runtime.
Similarly, the 24-qubit GS circuit runs in under 13 hours with the $[[31, 1, 7]]$ code.
Conversely, the $T/P_{succ}$ values become impractically large in the no-QEC cases, reaching $2.55 \times 10^{19}$ for the 2048-qubit QFT, which corresponds to approximately $8.10\times10^{6}$ years.
For the 24-qubit GS, $P_{succ}=0$ in all no-QEC cases, both for the optimal short chain and for the long chain with $C=100$.

In summary, incorporating the QEC code remains beneficial even when runtime is taken into account, with its advantages becoming increasingly pronounced for large-scale quantum algorithms.
The complete set of results for all benchmark algorithms for each simulation are provided in Sec. S1 of the Supplemental Material~\cite{suppl_mat}.

\section{Conclusion} 
\label{section: conclusion}

In this work, we propose a chip architecture for scalable, fault-tolerant quantum computing that is compatible with currently achievable ion-trap technologies.
By carefully arranging trap regions to support the qubit-connectivity requirements of logical gates and syndrome extraction, the proposed chip layout enables an efficient implementation of the two-dimensional color code and allows for a high level of parallelism in controlling physical components.

To evaluate the expected efficiency of the architecture, we also develop a software tool that simulates ion behavior during circuit execution and estimates system performance.
The tool models a real ion-trap system by incorporating necessary tasks such as transpilation and scheduling.
Our simulations on several benchmark quantum algorithms show that the proposed architecture significantly improves performance for large-scale quantum circuits.
With a physical two-qubit gate error probability of $1 \times 10^{-4}$, the architecture achieves an effective two-qubit gate error probability as low as $10^{-8}$ when encoding a logical qubit using the $[[31, 1, 7]]$ color code.
Although the use of QEC code introduces some runtime overhead, it is still advantageous as it yields significant improvements in the expected runtime for successful circuit execution.

In summary, this work provides a comprehensive design and demonstration of an ion-trap system architecture that paves the way from the NISQ regime to robust, fault-tolerant quantum computing.
We note that implementing large-scale systems will require additional procedures such as sympathetic cooling to mitigate motional heating caused by frequent ion shuttling.
Although this procedure introduces extra runtime overhead, it does not fundamentally alter the scalability advantages of the proposed architecture.
An extended scheduling policy incorporating sympathetic cooling, along with its performance implications, is discussed in detail in Sec. S2 of the Supplemental Material~\cite{suppl_mat}.
There also remains room for further optimization, including improved initial mapping of logical qubits to data ions and the development of more advanced scheduling policies.
Such future enhancements can be built upon the findings of this work and further accelerate the development of practical and scalable quantum computers.

\begin{acknowledgments}
This work was supported by Institute of Information \& Communications Technology Planning \& Evaluation (IITP), Grants No. RS-2022-II221040 and No. RS-2021-II211810, and the National Research Foundation of Korea (NRF) Grant No. RS-2024-00442855 and No. RS-2024-00413957, all funded by the Korean government (MSIT).

The data that support the findings of this article are openly available~\cite{Code}.
\end{acknowledgments}

\appendix
\section{Transpiler}
\label{appendix: transpiler}
The transpiler converts an input circuit into an equivalent or approximate circuit expressed entirely in native gates of the target hardware.
While most known quantum algorithms are described using a standard gate set (e.g., Clifford gates and non-Clifford gates like \textsc{toffoli} and $T$), each hardware platform, such as an ion-trap system, has its own set of directly executable gates.
For ion-trap systems, this native set typically includes single-qubit rotations around any axis on the Bloch sphere and the \textsc{ms} gate for two-qubit operations.
The transpiler's primary function is to decomposes each gate in the input circuit into an equivalent sequence of these native gates and to perform optimizations that reduce circuit depth and gate count.

In this work, we use Qiskit~\cite{qiskit24}, a widely-used framework for quantum circuit manipulation, as the foundation for our transpiler.
Initially, it conducts several optimizations at the standard gate set level.
In addition to Qiskit’s default optimization passes, we also implement custom passes that are not implemented by default, such as merging consecutive $Z$ axis rotations (e.g., sequences of $Z, S, S^{\dagger}, T, T^{\dagger}$) and canceling self-adjoint gate pairs, as well as reducing $H$ gates as described in Ref.~\cite{Nam18}.

After initial optimization, the circuit is decomposed into a basis set of $\{X, Y, Z, H, \textsc{cnot}, Rz\}$.
Because the resulting $Rz$ gates can have arbitrary angles, the transpiler generates two versions of the circuit: one having $Rz$ gates with an arbitrary angle and another with $Rz$-angle approximation for use in our fault-tolerant simulations.
The approximation represents arbitrary angle $\theta$ as a sum of $\theta_p$ [see Eq.~(\ref{eq: angle_decomposition})], where $\theta_p = 2\pi/2^p$ ($p\ge 3$ integer) and $t$ determines the precision, which is set to 16 in the simulation:

\begin{equation} \begin{split}
    \theta\ & \approx \ \sum_{i=1}^{t}{b_i \theta_i} =\sum_{i=1}^{t}{b_i \frac{2\pi}{2^i}} \quad \left(b_i \in \{0, 1\} \right) \\
    Rz(\theta)\ & \approx \ Rz(b_1 \theta_1) \circ Rz(b_2 \theta_2) \circ \cdots \circ Rz(b_t \theta_t)
    \label{eq: angle_decomposition}
\end{split}\end{equation}

\begin{figure}[ht]
    \includegraphics[width=6cm, trim=3.5 3.5 3.5 3.5]{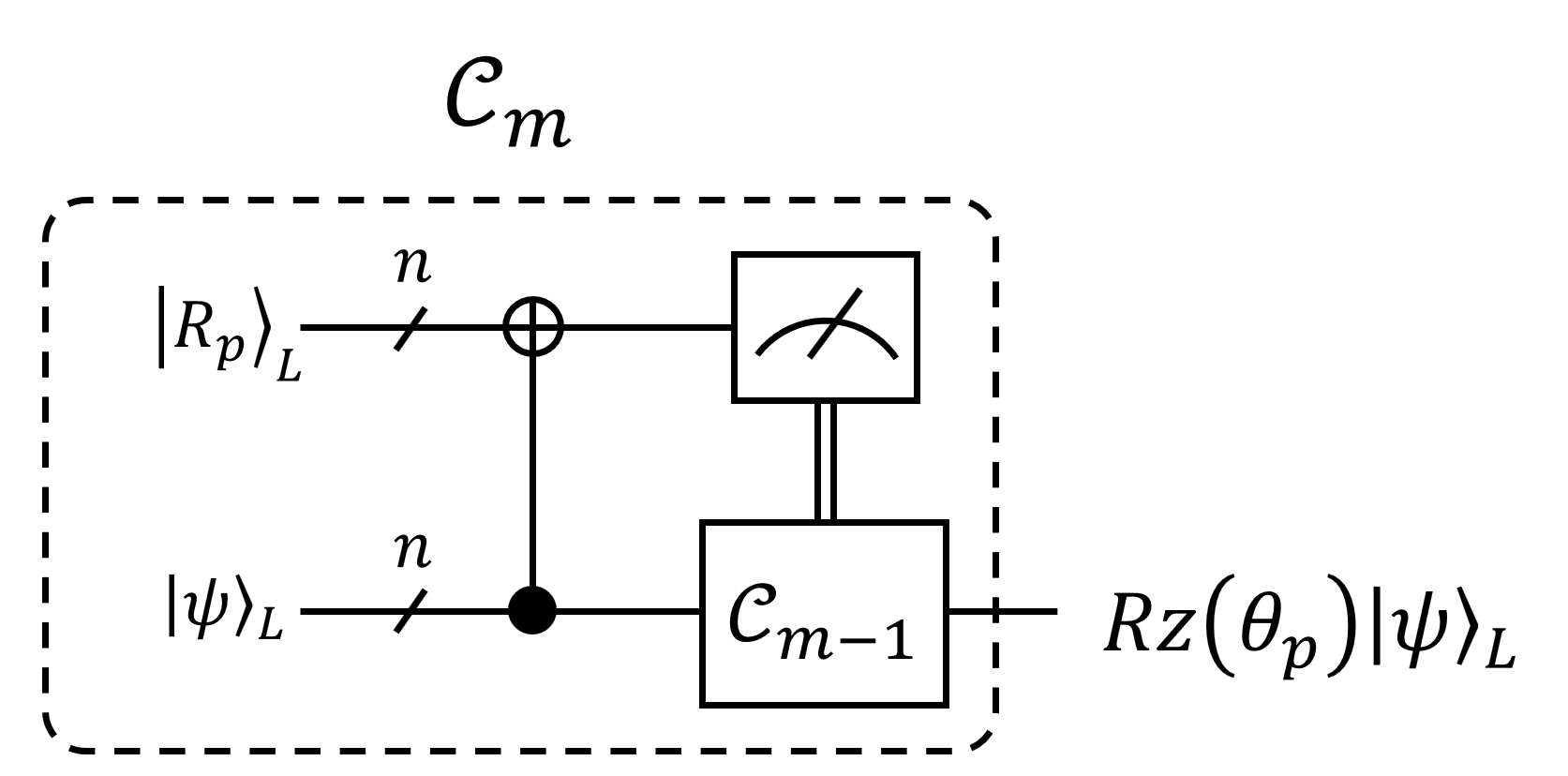}
    \caption{Quantum circuit to fault-tolerantly implement logical $Rz(\theta_p)$ gates in the higher Clifford hierarchy, given a logical state $|R_p\rangle_L$.}
    \label{fig: tgate_circuit}
\end{figure}

Each $Rz(\theta_p)$ gate is fault-tolerantly implemented by consuming a logical state $|R_p\rangle_L$ through a V sector (see Fig.~\ref{fig: tgate_circuit}).

Following these optimizations, the transpiler converts the circuit into a directed acyclic graph composed of native gates.
We adopt the native gate set of commercial ion-trap platforms, namely the \textsc{gpi}, \textsc{gpi2}, and \textsc{ms} gates \cite{IonqNative24}.
The matrix representations of these gates are given in Eqs.~(\ref{eq: native_1q_def}) and (\ref{eq: native_2q_def}):

\begin{equation} \begin{split}
    &\textsc{gpi}(\phi) = \begin{bmatrix} 0 && e^{-i\phi} \\ e^{i\phi} && 0 \end{bmatrix} \\ & \textsc{gpi2}(\phi) = \frac{1}{\sqrt{2}}\begin{bmatrix} 1 && -i e^{-i\phi} \\ -i e^{i\phi} && 1 \end{bmatrix} 
\end{split} \label{eq: native_1q_def} \end{equation}
\begin{equation}\begin{split}
    & \textsc{ms}(\phi_0, \phi_1) = \frac{1}{\sqrt{2}} \left[\begin{smallmatrix} 1 && 0 && 0 && -i e^{-i\Phi} \\ 0 && 1 && -i e^{-i\Delta \phi} && 0 \\ 0 && -i e^{i\Delta \phi} && 1 && 0 \\ -i e^{i\Phi} && 0 && 0 && 1 \end{smallmatrix}\right] \\
    & (\Phi \equiv \phi_0 + \phi_1,\ \Delta \phi \equiv \phi_0 - \phi_1)
\end{split} \label{eq: native_2q_def} \end{equation}

The native gate conversion rules are explicitly derived as in Eqs.~(\ref{eq: native_1q_conv_rule}) and (\ref{eq: native_2q_conv_rule}):

\begin{equation} \begin{split}
    &X = \textsc{gpi}(0) \ \ \ \ \ Y = \ \textsc{gpi}(\pi/2) \\
    &Z = i \cdot R_z(\pi) \ \ \ \ \ S = e^{i\pi/4} \cdot R_z\left( \pi/2 \right)\\
    &H = i \cdot R_z(\pi) \circ \textsc{gpi2}\left(3\pi/2 \right)
\end{split} \label{eq: native_1q_conv_rule} \end{equation}
\begin{equation} \begin{split}
    \textsc{cnot}_{1, 2} =& \exp{(-i\pi/4)}\cdot \textsc{gpi2}_1(-\pi/2) \circ \textsc{gpi2}_1(\pi) \\
    \circ \ & \textsc{gpi2}_2(\pi) \circ \textsc{ms}(0, 0, \pi/2) \circ \textsc{gpi2}_1(\pi/2)
    \label{eq: native_2q_conv_rule}
\end{split} \end{equation}

Here, $\textsc{cnot}_{1,2}$ denotes a \textsc{cnot} gate with qubit 1 as the control and qubit 2 as the target, and $\textsc{gpi}_i$ and $\textsc{gpi2}_i$ indicate the operation is applied to the qubit with index $i$.
For circuits without QEC, all $Rz$ gates can be deferred to the end of the circuit by adjusting the parameters of intermediate native gates, a process known as virtualization~\cite{McKay17}.
However, for circuits to be used with a QEC code, $Rz$ gates with $\pm \pi/4$ or smaller angle correspond to nontransversal gates that must be performed as special instructions in the V sector.
In this case, only $Rz(\pm \pi/2)$ and $Rz(\pi)$ gates (equal to $S, S^{\dagger}, Z$ respectively) are deferred.

After this conversion, the transpiler applies further native-level optimizations.
For example, two adjacent \textsc{gpi} gates can be merged into a single $Rz$ gate as shown in Eq.~(\ref{eq: native_opt_rule}).

\begin{equation}
    \textsc{gpi}(\theta_2) \circ \textsc{gpi}(\theta_1) = Rz\big(2(\theta_2 - \theta_1)\big)
    \label{eq: native_opt_rule}
\end{equation}
This optimization, followed by deferring the resulting $Rz$ gate, is repeatedly applied until no further simplification is possible.
As noted in Sec.~\ref{subsection: transpiler}, the correctness of the transpiler was verified on various circuits with 8–12 qubits by comparing the two state vectors produced by the original and transpiled circuits, respectively.

\section{Scheduling Policy}
\label{appendix: scheduling_policy}

In a real ion-trap system, a controller must map each qubit to a data ion and manage the execution of gate operations by applying lasers or adjusting trap voltages.
While each qubit has a unique next pending operation, there is flexibility in choosing which qubit's operation to execute first unless fully run in parallel.
However, a physical constraint exists: two qubits must be located in the same trap region to interact, such as for a two-qubit gate.
If they are in separate sectors, \textsc{swap} and shuttling operations must be inserted to bring the target qubits together.

The scheduler in our software tool handles all these tasks, including the initial qubit mapping, selecting the next gate operation, and dynamically inserting \textsc{swap} and shuttling operations.
The initial mapping assigns the $i$th qubit to the $i$th leftmost ion.
For subsequent scheduling, two heuristic policies are employed: the UB policy and the VTB policy.
The primary distinction between these policies lies in how they determine when to initiate a shuttling operation.

\subsection{UB Scheduling Policy}
\label{appendix: ub_policy}
The UB scheduling policy greedily executes all possible gate operations that can be performed before shuttling, and only inserts a shuttling operation when no further operations are possible with the current ion configuration.
This situation occurs if, for each qubit in a sector, one of the following conditions is met:

\begin{itemize}
    \item The next gate is a two-qubit gate whose pair qubit is in a different H sector.
    \item The next gate is blocked by a prerequisite of the pair qubit (e.g., several gate operations performed on the pair qubit must precede it).
    \item The next gate is a nontransversal operation.
    \item The qubit has no remaining gates because all its operations have been completed.
\end{itemize}
Meanwhile, all qubits in the V sectors perform either syndrome extraction or a nontransversal gate followed by syndrome extraction.

Once no further operations are possible, the ions are shuttled following the rules in Sec.~\ref{section: chip_architecture}.
During shuttling, a logical qubit in a V sector enters an adjacent H sector, while a new qubit enters the V sector.
The UB policy ensures that the qubit with the highest priority is positioned at the end of its sector before shuttling so it can be moved to the V sector.
Priority is determined by marking qubits whose pending gate is nontransversal or a two-qubit operation that requires a pair to move closer.
\begin{algorithm}[H]
    \caption{UB Scheduling Policy}
    \label{alg: ub_policy}
    \begin{algorithmic}[1]
    \State $F$ : List of all pending gates of qubits
    \State $L$ : List of gates which are not executable in current configuration
    \medskip
    \State Initialize $F$ with the first pending gate of each qubit
    \While{$F$ is not empty}
        \State $gate\ \leftarrow \ F.pop()$
        \If{$gate$ is executable}
            \State Schedule $gate$
            \For{$child$ in $gate.successor$}
                \State $F.append(child)$
            \EndFor
        \Else
            \State $L.append(gate)$
        \EndIf
    \EndWhile
    \For{$gate$ in $L$}
        \If {$gate$ is nontransversal}
            \State Mark $gate.qubit$ as movable
        \ElsIf{$gate$ is two-qubit gate}
            \State /* two qubits of $gate$ are in different sector */
            \If{$right\_phase$}
                \State Mark left qubit of $gate$ as movable
                \State Mark right qubit of $gate$ as immovable
            \Else
                \State Mark right qubit of $gate$ as movable
                \State Mark left qubit of $gate$ as immovable
            \EndIf
        \EndIf
    \EndFor

    \For{each $sector$ in the chip}
        \State /* $sq$ : swapped qubit */
        \If{$right\_phase$}
            \If{movable qubit exists}
                \State $sq\ \leftarrow$ rightmost movable qubit within $sector$
            \Else
                \State $sq \ \leftarrow$ rightmost not immovable qubit within $sector$
            \EndIf
            \If{$sq \ne$ rightmost qubit of $sector$}
                \State Schedule \textsc{swap} (rightmost qubit, $sq$)
            \EndIf
        \Else
            \If{movable qubit exists}
                \State $sq\ \leftarrow$ leftmost movable qubit within $sector$
            \Else
                \State $sq \ \leftarrow$ leftmost not immovable qubit within $sector$
            \EndIf
            \If{$sq \ne$ leftmost qubit of $sector$}
                \State Schedule \textsc{swap} (leftmost qubit, $sq$)
            \EndIf
        \EndIf
    \EndFor
    \medskip
    \State Apply QEC to all qubits in V sector
    \If{$right\_phase$}
        \State Schedule shuttling (right)
    \Else
        \State Schedule shuttling (left)
    \EndIf

    \If{rightmost H sector is full}
        \State $right\_phase \leftarrow$ False
    \ElsIf{leftmost H sector is full}
        \State $right\_phase \leftarrow$ True
    \EndIf
    \end{algorithmic}
\end{algorithm}

\begin{algorithm}[H]
    \caption{VTB Scheduling Policy}
    \begin{algorithmic}[1]
    \State $max\_runtime\_bound = 0$
    \For{each $V\_sector$ in the chip}
        \State $runtime\_bound = 0$
        \State /* Only one $qubit$ in the $V\_sector$ */
        \State $qubit \ \leftarrow \ V\_sector.qubit$
        \If{$qubit.next\_gate$ is nontransversal}
            \State Schedule $qubit.next\_gate$
            \State $runtime\_bound\ \mathrel{+}=\ execution\_time$
        \EndIf
        \State Schedule QEC operation to $qubit$
        \State $runtime\_bound\ \mathrel{+}=\ QEC\_time$
        \If {$runtime\_bound > max\_runtime\_bound$}
            \State $max\_runtime\_bound \leftarrow runtime\_bound$
        \EndIf
    \EndFor
    \For{each $H\_sector$ in the chip}
        \State $trap\_runtime = 0$
        \If{$right\_phase$}
            \State $order \leftarrow$ list of qubits from right to left
        \Else
            \State $order \leftarrow$ list of qubits from left to right
        \EndIf
        \For{$qubit$ in $order$}
            \If{$qubit.next\_gate$ is executable}
                \State $trap\_runtime \mathrel{+}=qubit.next\_gate.runtime$
                \If{$trap\_runtime \le max\_runtime\_bound$}
                    \State Schedule $qubit.next\_gate$
                \Else
                    \State $trap\_runtime\mathrel{-}=qubit.next\_gate.runtime$
                    \State Break loop
                \EndIf
            \EndIf 
        \EndFor
        \If{$max\_runtime\_bound-trap\_runtime \ge 
        \text{SWAP}\_time$}
            \State $sq \leftarrow$ Qubit to be shuttled
            \If{$right\_phase$ and $sq \ne$ rightmost qubit of $H_sector$}
                \State Schedule \textsc{swap} (rightmost qubit, $sq$)
            \ElsIf{not $right\_phase$ and $sq \ne$ leftmost qubit of $H_sector$}
                \State Schedule \textsc{swap} (leftmost qubit, $sq$)
            \EndIf
        \EndIf
        \If{$right\_phase$}
            \State Schedule shuttling (right)
        \Else
            \State Schedule shuttling (left)
        \EndIf
        \If{rightmost H sector is full}
            \State $right\_phase \leftarrow \text{False}$
        \ElsIf{leftmost H sector is full}
            \State $right\_phase \leftarrow \text{True}$
        \EndIf
        \EndFor
    \end{algorithmic}
    \label{alg: vtb_policy}
\end{algorithm}

The scheduler then selects the rightmost marked qubit during a right-shuttling phase and the leftmost marked qubit during a left-shuttling phase to be moved out of the current H sector.
If the selected qubit is not already at the end of the H sector, a \textsc{swap} operation is inserted to move it to the end of the H sector.
This process of executing all possible gates, inserting \textsc{swap} operations, and shuttling is repeated until all gate operations are exhausted. 
The full procedure is detailed in Algorithm~\ref{alg: ub_policy}.

\subsection{VTB Scheduling Policy}
\label{appendix: vtb_policy}
In contrast to the UB policy, the VTB policy initiates a shuttling operation immediately after all operations for qubits in the V sectors are completed.
For a qubit in the V sector, its operation time is the sum of its (nontransversal) gate execution time and the time for syndrome extraction.
Formally, the shuttling period $\tau$ is defined by the maximum time required to complete pending operations across all V sectors, as shown in Eq.~(\ref{eq: vtb_period}):

\begin{equation}
    \tau = \max_{i}{\left( t_{\textsc{nt}, i}+t_{\textsc{synd}} \right)}
    \label{eq: vtb_period}
\end{equation}

Here, $t_{\textsc{nt}, i}$ is the time for a nontransversal gate if the pending gate is nontransversal, and 0 otherwise, while $t_{\textsc{synd}}$ is the time for syndrome extraction.
Other qubits in H sectors must complete their (transversal) gate operations within this period $\tau$.
As expressed in Eq.~(\ref{eq: vtb_runtime}), the sum of the times for all single-qubit, two-qubit, and \textsc{swap} operations in any H sector must not exceed $\tau$.
Formally, let $n_{1Q, j}$, $n_{2Q, j}$, and $n_{\textsc{swap}, j}$ denote the number of single-qubit gates, two-qubit gates, and \textsc{swap} operations, respectively, performed between two shuttling operations in the $j$th sector.
Similarly, let $t_{1Q}$, $t_{2Q}$, and $t_{\textsc{swap}}$ be the corresponding gate times, as listed in Table~\ref{tab: gate_info}.
Then the VTB policy must satisfy Eq.~(\ref{eq: vtb_runtime}).

\begin{equation}
    n_{1Q, j}\cdot t_{1Q} + n_{2Q, j}\cdot t_{2Q} + n_{\textsc{swap}, j}\cdot t_{\textsc{swap}}\ \le \ \tau
    \label{eq: vtb_runtime}
\end{equation}

Within each interval $\tau$, the VTB scheduler prioritizes gates of qubits that will be shuttled next.
For example, in a right-shuttling phase, it first executes all pending gates of the rightmost qubit.
If time remains, it proceeds to the second rightmost qubit, and so on, until the period $\tau$ is exhausted or no further gates can be executed.
The selection of qubits to be swapped follows the same rules as the UB policy, provided enough time for the \textsc{swap} operation remains.
The complete VTB scheduling algorithm is detailed in Algorithm~\ref{alg: vtb_policy}.

The more frequent shuttling in VTB scheduling policy can lengthen the total runtime, but it also provides more opportunities for error correction and the execution of nontransversal gates.
This creates a complex trade-off between runtime and success rate as clearly observed in the results of Sec.~\ref{section: benchmark}, where VTB typically shows a better success rate and longer runtime than the UB policy.

\nocite{*}

\bibliography{000_references}

\clearpage
\foreach \x in {1,...,7}
{
\clearpage
\includepdf[pages={\x}]{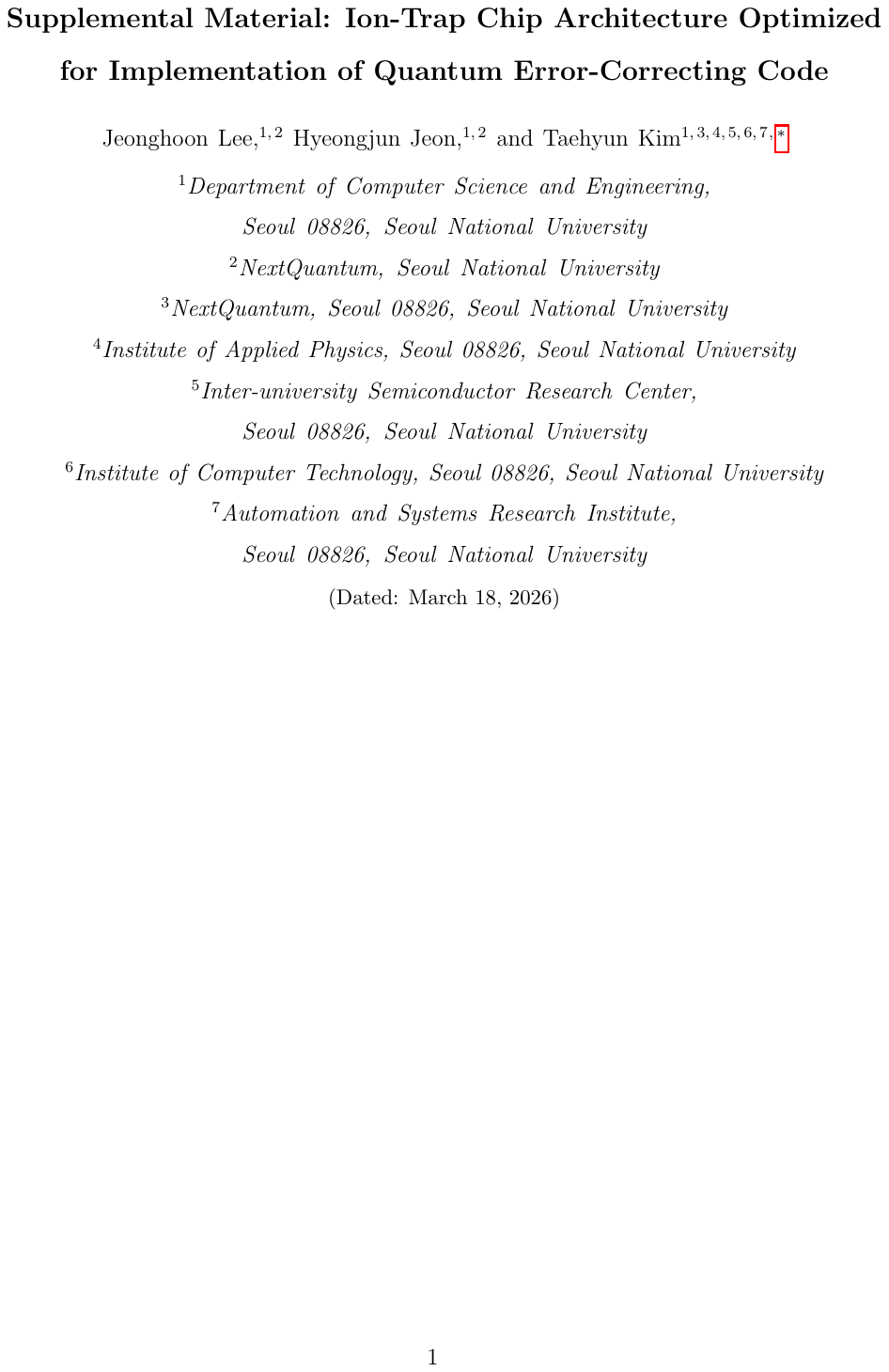}
}

\end{document}